# Multi-Objective Bayesian Materials Discovery:
## Application on the Discovery of Precipitation Strengthened NiTi Shape Memory Alloys through Micromechanical Modeling

Alexandros Solomou[a], Guang Zhao[b], Shahin Boluki[b], Jobin K. Joy[a], Xiaoning Qian[b], Ibrahim Karaman[c], Raymundo Arróyave[c], Dimitris C. Lagoudas[a,c]

[a]Department of Aerospace Engineering, Texas A&M University, College Station, TX 77843
[b]Department of Electrical & Computer Engineering, Texas A&M University, College Station, TX 77843
[c]Department of Materials Science and Engineering, Texas A&M University, College Station, TX 77843

**ABSTRACT**

In this study, a framework for the multi-objective materials discovery based on Bayesian approaches is developed. The capabilities of the framework are demonstrated on an example case related to the discovery of precipitation strengthened NiTi shape memory alloys with up to three desired properties. In the presented case the framework is used to carry out an efficient search of the shape memory alloys with desired properties while minimizing the required number of computational experiments. The developed scheme features a Bayesian optimal experimental design process that operates in a closed loop. A Gaussian process regression model is utilized in the framework to emulate the response and uncertainty of the physical/computational data while the sequential exploration of the materials design space is carried out by using an optimal policy based on the expected hyper-volume improvement acquisition function. This scalar metric provides a measure of the utility of querying the materials design space at different locations, irrespective of the number of objectives in the performed task. The framework is deployed for the determination of the composition and microstructure of precipitation-strengthened NiTi shape memory alloys with desired properties, while the materials response as a function of microstructure is determined through a thermodynamically-consistent micromechanical model.

## 1. INTRODUCTION

Recent advances in manufacturing and computational resources have led to sophisticated designs of new devices suitable for aerospace, automotive and biomedical applications, among others. The functionality of such devices is based on the exploitation of the capabilities of multifunctional materials, and therefore, their development has motivated the search for novel materials with optimal behavior.

The discovery of new materials is bound by the resource-intensive nature of the materials discovery process [1]–[3]. Intuition-based approaches are limited in that they only allow the investigation of a very small fraction of a given materials design space which in turn typically consists of a large number of degrees of freedom, including, but not limited to, microstructural, structural and chemical material attributes [4]. Typical approaches to tackle this problem include the utilization of High-Throughput (HT) computational [5] and experimental frameworks [6], [7], which are used to generate large databases of materials feature / response sets, which then must be analyzed [8] to identify the materials with the desired characteristics.

HT methods, however, do not account for constraints in (experimental / computational) resources available, nor account for the existence of bottle necks in the scientific workflow that necessarily prevent the parallel execution of specific experimental / computational tasks. As an example of the latter limitation, one can imagine experimental programs where the HT synthesis of materials libraries is ultimately throttled by the low-throughput nature of necessary materials characterization operations.

Very recently, global optimization frameworks—including gradient based [9], [10], direct search methods [11], [12] and Bayesian optimization approaches [13]–[19] have emerged to guide the efficient exploration of the material space [4], [20]–[29]. Unlike high throughput-based approaches to materials discovery, Bayesian Optimization (BO) enables the (global) optimization of materials while minimizing the number of evaluations of the materials space. This is realized by sequentially executing policies in a way that ensures the balance between the exploitation of the design space and its exploration [12]. The applicability of these techniques has been successfully demonstrated in a few materials science problems [4], [20]–[30], although the published work tends to focus on the optimization of a single objective (such as NiTi-based SMAs with very low thermal hysteresis [20]). To the best of our knowledge, however, there has been little work [29], [30] focused on developing optimal policies for sequential experimental design in which multiple materials attributes / performance metrics must be optimized at once. As described below, the present work focuses on the use of Bayesian global optimization techniques to perform Multi-Objective (up to 3 objectives) Bayesian Optimal Experimental Design (BOED) and to guide the sequential query of the materials design space for the discovery of materials with desired properties. In particular, the framework is deployed to discover the optimal composition and microstructural features (i.e. the precipitate volume fraction) of precipitation-strengthened NiTi shape memory alloys (SMAs).

BOED methods typically operate in an iterative loop that includes three major steps: (1) the machine-learning step, (2) the selector step, and (3) the database (information)-updating step. During the machine-learning step, a probabilistic surrogate model is constructed to model the functional relationships between the inputs and outputs of the system, *accounting for the uncertainty in the system's response*. When a material system is considered, the input variables are the selected degrees of freedom (chemical, microstructural features or even process parameter conditions) that modify the material's response. The performance/property of the system is the output variable, which is a function of the (controllable) materials degrees of freedom. A surrogate model is then constructed based on available data encompassing the values of the measured input-output variables. The aforementioned data can potentially include data measured from both physical and computational experiments.

During the selector step, an acquisition function that explicitly accounts for the uncertainty in the probabilistic surrogate model predictions is used to determine the next point in the materials design space to evaluate. As the acquisition function characterizes the expected utility of each point in the material's design space – based on its predicted performance with uncertainty by the probabilistic surrogate model – the trade-off between "exploitation" and "exploration" can be balanced by the proper design of this acquisition function [14], [23]. An acquisition function emphasizing "exploitation" biases the search towards the optimality of the desired performance under the current surrogate model while an acquisition function emphasizing "exploration" aims to optimally reduce the uncertainty of the surrogate model [23].

Once the new materials degrees of freedom (input parameters) to test is identified, during the database (information)-updating step, a new set of (physical or computational) experiments is conducted so that the property / performance-related parameters of the system under study are calculated and the database of the available input-output data is updated. Upon the completion of this step, a new iteration of the BOED framework begins.

There are many Bayesian global optimization techniques with different policies/acquisition functions that are widely used to perform BOED such as the Efficient Global Optimization (EGO) [31], which is based on the Expected Improvement (EI) acquisition function, the Knowledge Gradient (KG) [17], and the BOED based on the Mean Objective Cost of Uncertainty (MOCU) [16], [22], among others. For instance, Seko *et al.* [27] employed a BO framework to discover compounds of ordered crystal structures with the highest melting point. To accomplish this, they utilized the Probability of Improvement (PI) acquisition function in the selector step. Furthermore, they evaluated the performance of various probabilistic surrogate models including the Gaussian Process Regression (GPR), least-squares regression (OLSR), partial least-squares regression (PLSR) and support vector regression (SVR). Balachandran *et al* [23] applied a BO with a GPR model and the EI acquisition function on the design of $M_2AX$ phases with of maximum elastic moduli. Dehghannasiri *et al* [22] proposed MOCU-based BOED to identify SMAs with the lowest energy dissipation at a specific temperature. Ueno *et al* [26] designed a BO framework to determine the atomic structure of a crystalline interface by using Thompson sampling and Bayesian linear regression combined with a random feature map (that approximates a GPR model) to model the response of the materials design space. In later work, Seko *et al.* [25] employed BO based on the GPR model and PI acquisition function to discover low thermal conductivity compounds. While most approaches have been implemented over a computational space, Xue *et al.* [20], [21] used BO to accelerate *the experimental* and computational discovery of NiTi-based SMAs ($Ti_{50.0}Ni_{46.7}Cu_{0.8}Fe_{2.3}Pd_{0.2}$) with very low thermal hysteresis and $BaTiO_3$-based piezoelectrics with vertical morphotropic phase boundary. In [20], Xue *et al.* performed BOED using the EGO framework and KG while utilizing various probabilistic models (GPR, Support Vector Regression (SVR) with a radial basis function kernel and with a linear kernel and using bootstrap uncertainty estimates). In [21] they have employed a Bayesian linear regression model and incorporated some constraints from domain knowledge via a truncated Gaussian prior along with the greedy approach (pure exploitation). More recently, Ju *et al.* [28] put forward a computational closed-loop framework based on the Bayesian regression model of [26] and the EI acquisition function for the efficient discovery of Si-Ge composite interfacial structures that minimize or maximize the interfacial thermal conductance across Si-Si and Si-Ge interfaces. Prior approaches, with some exceptions [29], [30] have focused on single-objective materials optimization/discovery problems.

The current work utilizes an BOED approach based on the EHVI [15] acquisition function to perform multi-objective optimization. EHVI balances the trade-off between exploration and exploitation for multi-objective BOED problems, similar to EI for single-objective problems. Here we note that EHVI is a scalar quantity that allows a rational agent to select, sequentially, the next best experiment to carry out, given current knowledge, *regardless of the number of objectives*, or dimensionality, of the materials discovery/development exercise. Finally, in the current work a GPR model [32] is used as the probabilistic surrogate model to capture the non-linear functional relationships of the input-output variables for the material system under study.

In the present study, we demonstrate the applicability of the BOED framework on the discovery of precipitation strengthened NiTi SMAs for selected target properties. SMAs are materials that have the capability of converting thermal energy into mechanical work and the ability to generate and recover moderate to large inelastic deformations, which is made possible through a diffusionless and reversible solid-to-solid martensitic phase transformation and makes them suitable for various applications ranging from actuators [33]–[37] to applications where high flexibility (i.e. superelasticity) is required such as medical implants and instruments [38], [39].

Experimental studies in aged NiTi SMAs have shown that the presence of (metastable) $Ni_4Ti_3$ significantly influences its phase transformation characteristics and properties, such as transformation hysteresis (which dictates the efficiency of energy conversion), transformation temperatures and transformation strains [40]–[43] and enhances the stability of cyclic actuation and superelastic responses, and improves fatigue lives [44]–[46]. Furthermore, the properties of NiTi SMAs are extremely sensitive to the compositional variations. For example, 1 % at. change in Ni content above 50 % at. Ni can reduce the transformation temperatures about 80°C and transformation hysteresis about 20°C (40% reduction). More importantly, the formation of metastable (metastable) $Ni_4Ti_3$ precipitates not only brings about structural heterogeneities in the microstructure but also leads to the appreciable local compositional changes, affecting the martensitic transformation characteristics, properties (such as the extent of the transformation range), and performance [40], [47]–[50]. Because of the complications arising from extreme sensitivity of the properties to the composition and microstructural heterogeneities in NiTi SMAs, the design and selection of proper NiTi SMAs, for specific applications in biomedical devices, aerospace, and energy related fields (especially in oil and gas applications), with desired properties (in particular transformation temperatures and hysteresis, and stress-required for martensitic transformation) have been cumbersome, time consuming, and often a limiting factor for the insertion of these materials into some of these application areas. Therefore, an intelligent choice of the composition and thermal processing which leads to desired precipitate distributions in precipitation hardened SMAs can be a key factor to acquire materials that meet the property requirements of a targeted application and help their accelerated insertion.

In this work, rather than focusing on the exploration of the entire process-structure-performance (PSP) space, we focus on the simpler problem of efficiently determining the microstructural characteristics, represented in this context as matrix composition and volume fraction of precipitates *that yield desired properties* in NiTi-based SMAs. To this end, a previously validated SMA micromechanical model [51]–[53] which links matrix-precipitate microstructures with the effective material response is used to explore the influence of precipitation on the thermo-mechanical response of NiTi SMAs.

The high fidelity SMA model is used in the BOED framework to conduct computational experiments and augment the data used in the machine-learning step to construct an updated surrogate model. It is worth pointing out that the decision to use the SMA micromechanical model was based on the need to demonstrate the BOED framework under controlled circumstances, with full knowledge of the (computational) ground truth. These factors contribute to the systematic, thorough evaluation of the proposed methodology.

Moreover, and *more importantly*, the focus on computational models that mimic the experimental efforts enables the implementation of a full closed-loop approach in which the experimental design is carried out autonomously by the computational algorithm, without human input. The major contribution of the present work is the use of a scalar metric (EHVI) in the sequential querying of materials design space in which multiple objectives (up to 3 in the present work) are to be achieved at once.

The remainder of the paper is organized as follows. In Section 2, the formulation of the SMA micromechanical model is presented. The model is calibrated based on the response of NiTi SMAs without precipitates (solution heat treated case) and the validity of its results on predicting the responses of both solution heat treated (SHT) and precipitation strengthened SMAs is demonstrated through correlations with experimental results. In Section 3, the overview of the BOED framework is described while the mathematical background of the used GPR model and the EHVI acquisition function is provided. In Section 4, the capabilities of the developed BOED framework are demonstrated by solving two distinct materials discovery problems where precipitations hardened SMAs with properties that satisfy 2 and 3 objectives, respectively are queried. The solved problems suggest SMAs that meet the property requirements for specific aerospace applications where they can be used as solid-state actuators. Furthermore, the results also quantify the efficiency of the BOED framework in finding the materials with desired properties in comparison to the HT approaches. In addition, the utility of the queried materials by the BOED framework within a predefined experimental budget is compared with the utility of the corresponding queried materials following a Pure Random Experiment Selection (PRES) policy and a Pure Exploitation Experiment Selection (PEES) policy. Finally, in the last section the conclusions of the current work are summarized.

## 2. SMA MICROMECHANICAL MODEL

### 2.1 Modeling Framework

The adopted model captures the effect of the thermo-mechanical loading conditions (see dashed or continues lines marked with different colors in Figure 1) and the microstructure of the material (see same color lines in Figure 1) on the effective response of the precipitated NiTi SMAs.

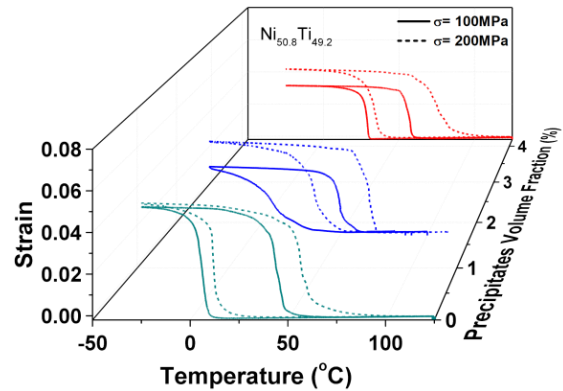

**Figure 1.** Experimentally observed tensile strain - temperature response of the $Ni_{50.8}Ti_{49.2}$ SMAs with different precipitate volume fractions for a thermal cycle under constant uniaxial tensile stress $(\sigma)$ conditions [51]. Different precipitate volume fractions is a consequence of different heat-treatment conditions.

The model, will be used in the developed BOED framework, which is presented in Section 3, in order to replicate the physical experiments and to enable the discovery of precipitation strengthened SMAs with desired properties.

A Finite Element (FE) based micromechanical modeling approach is adopted that uses Representative Volume

Element (RVE) models for statistically homogeneous materials to extract information on the effective response of precipitation strengthened NiTi SMAs. Methodologies considering only the phase-averaged response of the various constituents are defined to be mean-field approaches while full-field methods are those in which the position dependent field values are determined and then averaged for the effective, macroscopic response. Here, we use an FE based micromechanical modeling full-field approach which has been recently developed and validated by the authors [51]–[53] and has the ability to capture the effective macroscopic response of SMA materials by taking into consideration details in the microstructure which cannot be fully captured by mean field methods [54], [55] (e.g. Ni depletion in Ni-rich NiTi SMA upon precipitation). The developed method considers the SMA material as a statistically homogeneous material and analyzes its response based on RVE models. The RVE model, is a sufficient large subvolume of a statistically homogeneous microstructure which contains the same microstructural information (i.e. same phase volume fractions and statistical distributions) as the material-at-large and exhibits the same effective thermo-mechanical response. In the present work, the developed RVE model of the precipitation strengthened NiTi SMAs is generated on the basis of the input values of the materials initial homogeneous Ni concentration before precipitation ($c$) and the resulting precipitate volume fraction ($v_f$) after aging heat treatments. The model takes into consideration the chemical changes in the microstructure of the material due to the Ni depletion from the SMA matrix, associated with $Ni_4Ti_3$ precipitate formation [47], as well as the structural changes associated with the precipitation process to capture the thermo-mechanical response of the precipitation strengthened material depending on the applied thermo-mechanical loading conditions.

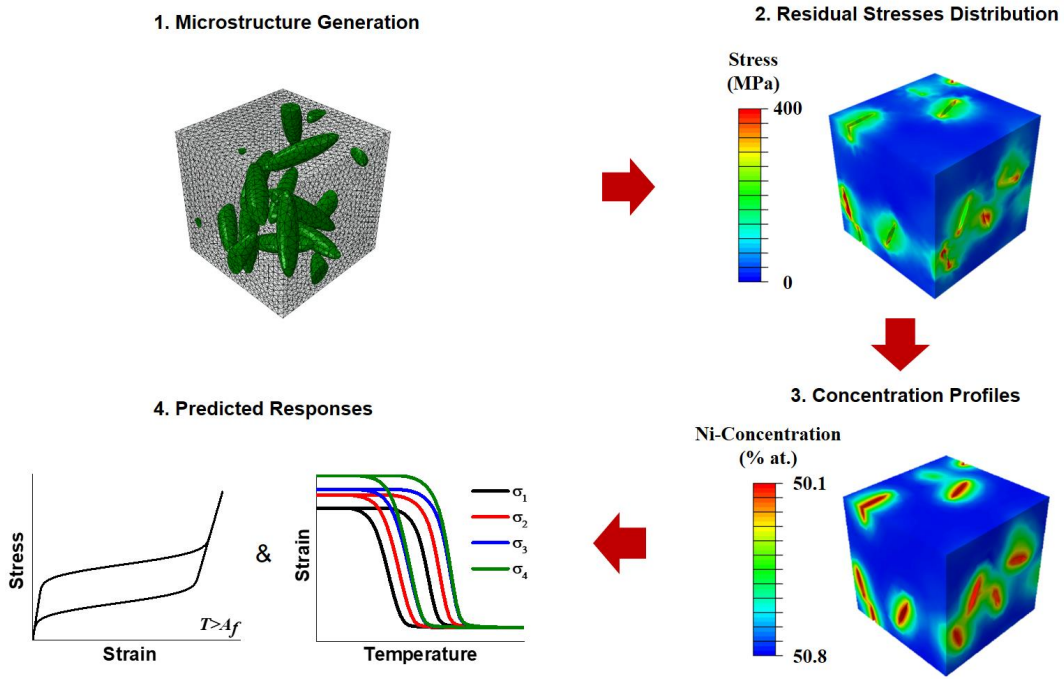

**Figure 2.** Micromechanical model framework for precipitation strengthened SMAs.

The workflow followed in the current study to generate the RVE model of the precipitation strengthened NiTi SMAs and to acquire the materials response is summarized below; the reader is refered to the references [51]–[53] for in-depth discussions related to the development and implementation of the RVE model. As shown in Figure 2, during the first step, the geometry of the microstructure of the RVE is constructed by generating 15 randomly distributed non-overlapping precipitates.

The number of the precipitates is selected such that the generated RVE model is considered as representative of the material-at-large [52] while their volume fraction is defined based on the given $v_f$ value which is provided as input to the RVE model. Furthermore, the shape of the precipitates for the considered NiTi material is assumed to be oblate spheroid with a major-to-minor axis ratio ~4:1 [51] while periodic boundary and geometry conditions are maintained to ensure complete periodicity on the developed RVE model. Details on the choice of the selected boundary and geometry conditions can be found in [51]–[53].

During the precipitation of the metastable Ni$_4$Ti$_3$ particles, Ni depletion in the matrix close to the precipitates creates non-uniform Ni concentration distributions in the surrounding matrix, which significantly affects materials transformation temperatures [56]. To capture this effect, during the second step of the process, the Ni concentration in the SMA matrix after precipitation is calculated using the Fickian diffusion law,

$$\dot{c}_A = D_A \nabla^2 c_A, \qquad (1)$$

which is solved using the FE method on the RVE domain. In equation (1), $c_A$ denotes the Ni concentration in the SMA matrix, the dot denotes differentiation with respect to time, and $D_A$ is the diffusivity coefficient. To this end, to solve the diffusion problem, periodic boundary conditions at the faces of the RVE are applied, the initial Ni concentration of the matrix material is set equal to the $c$ input value while on the precipitates boundaries a constant Ni concentration is set, equal to the matrix composition in equilibrium with the precipitates ($c_{eq}$). Based on the aforementioned initial and boundary conditions the equation (1) is solved until the average Ni concentration in the SMA matrix ($\bar{c}_A$) is equal to the value that is calculated by [51]–[53],

$$c = v_f c_{Ni_4Ti_3} + (1 - v_f)\bar{c}_A. \qquad (2)$$

In equation (2), $c_{Ni_4Ti_3}$ is the Ni concentration of the Ni$_4$Ti$_3$ precipitates, i.e. $c_{Ni_4Ti_3} = 56.8$ % at. $c_{eq} = 50.1$ % at. is calculated through the NiTi phase diagram reported in [40] while the diffusivity coefficient is considered as $D_A = 1.0845 \times 10^{-15}\,\text{m}^2\text{s}^{-1}$ [57]. In the same step, the phase transformation temperatures are assigned at each material point depending on their Ni concentration in accordance with the experimental results, for homogeneous NiTi SMAs (solution heat treated, no Ni$_4$Ti$_3$ precipitates) with different Ni concentrations. Differential Scanning Calorimetry (DSC) experimental data that link the transformation temperatures with the Ni concentration of homogenized NiTi SMAs are reported in [56] (Figure 3) and are used in the present work.

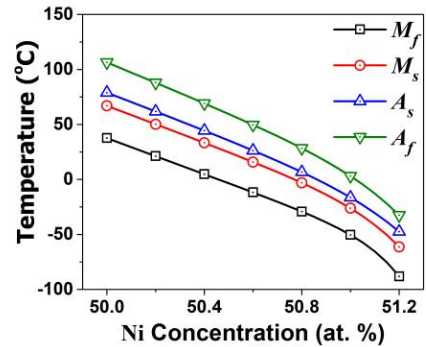

**Figure 3.** Effect of Ni concentration of homogeneous NiTi SMAs on the transformation temperatures [56].

In the third and fourth steps of the framework, the mechanical equilibrium equations are considered,

$$\nabla \cdot \boldsymbol{\sigma} + \mathbf{f} = 0, \quad (3)$$

and solved, using the FE method on the RVE domain, in the absence of body forces ($\mathbf{f}$) in order to capture the structural changes in the microstructure associated with the precipitation process and the thermo-mechanical response of the precipitation strengthened material depending on the applied thermo-mechanical loading conditions. Note in equation (3), $\boldsymbol{\sigma}$ denotes the Cauchy second order stress tensor, while in the current work it is assumed that the heating–cooling loading rates are sufficiently slow to justify the assumption of constant temperature throughout the RVE. To this end during the third step of the framework, the residual stresses due to the lattice mismatch of precipitates and SMA matrix are modeled by introducing eigenstrains into the precipitates. The values of the imposed eigenstrains,

$$\varepsilon_{ij}^{Ni_4Ti_3} = \begin{bmatrix} -0.00417 & 0 & 0 \\ 0 & -0.00417 & 0 \\ 0 & 0 & -0.00257 \end{bmatrix}, \quad (4)$$

were reported in [58], which are calculated using the lattice constants of the austenitic-B2 and $Ni_4Ti_3$-rhombohedral phases determined from the x-ray diffraction data. At the completion of the third step the RVE of the precipitated material is complete and during the fourth step of the framework, the generated RVEs are subjected to the desired thermo-mechanical loads to get the effective response of the precipitated material.

In the developed RVE, to model the constitutive response of the SMA matrix a widely accepted and validated thermodynamically consistent constitutive model of polycrystalline SMAs, developed by Lagoudas *et al.* [59], is used while for the precipitates, a linear elastic isotropic behavior is assumed. The used elastic properties of the preciptiates were derived from the first-principles calculations [60], while the material parameters needed in the SMA constitutive model are reported in the earlier works of the authors [52] and determined by performing uniaxial experiments, on solution heat treated $Ni_{50.8}Ti_{49.2}$ SMA samples, following the material characterization procedures described in [59], [61]. The values of the material parameters of the NiTi SMA matrix and the $Ni_4Ti_3$ precipitates used in the current work, are summarized in Table 1. It should be noted that size effects, i.e. the role of interparticle distance on the nucleation of martensite, are not captured in the current work, which can be quite important.

**Table 1.** Material parameters

| Solution heat treated $Ni_{50.8}Ti_{49.2}$ SMA parameters [52] | | $Ni_4Ti_3$ precipitate parameters [60] | |
|---|---|---|---|
| Material Parameter | Value | Material Parameter | Value |
| $E_A$ | 68 GPa | $E$ | 104 GPa |
| $E_M$ | 43 GPa | $v$ | 0.39 |
| $v_A = v_M$ | 0.33 | | |
| $H_{sat}$ | 0.051 | | |
| $k_t$ | 0.05 MPa$^{-1}$ | | |

| | | |
|---|---|---|
| $C_A$ | 7 MPa/K | |
| $C_M$ | 7 MPa/K | |

especially for the very small precipitates and high $c$ values [62]. Finally, to estimate the effective thermo-mechanical response of the RVEs, the volume average stress and total strain over the RVE, are calculated.

In summary, provided the values of the necessary SMA micromechanical model parameters, the model can be used to simulate the response of precipitation strengthened NiTi SMAs of different values of $c$ and $v_f$ under the desired thermo-mechanical loading conditions in lieu of experiments.

In the current work, the presented micromechanical model is incorporated in the developed BOED framework, which guides the sequential search of the SMA design space— the values of the $c$ and $v_f$ design parameters— to determine materials with targeted properties. Then the SMA micromechanical model for the given values of the design variables is used to perform computational experiments and to predict the response of the precipitation strengthened material (Figure 2). The results of the computational experiments are used to calculate the material parameters of a corresponding *homogenized* SMA – see Appendix A for a brief description of the calculation of the behavior of the *homogenized* SMA [59], [61]. Finally using the calculated material properties, the values of the selected objective functions are calculated and are provided to the BOED framework in order to proceed and select the next material to test.

## 2.2 Model Validation

A comparison between the predicted response of homogeneous $Ni_{50.8}Ti_{49.2}$ SMAs by the calibrated model and the corresponding experimental data under isobaric loading paths is presented in Figure 4. The solid curves represent experimental data at given constant uniaxial tensile stress conditions while the dashed ones represent the simulations.

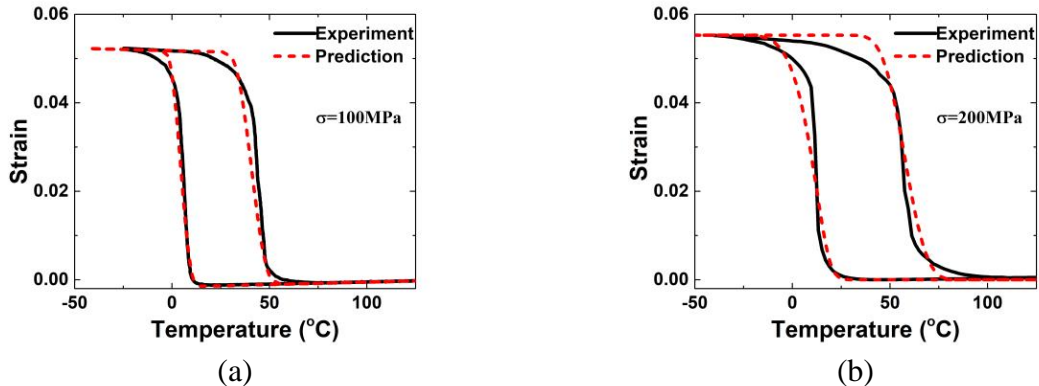

(a)            (b)

**Figure 4.** Predicted and experimentally measured tensile strain - temperature response of the homogeneous $Ni_{50.8}Ti_{49.2}$ SMAs for a thermal cycle under constant uniaxial tensile stress ($\sigma$) conditions [51]: (a) $\sigma = 100$ MPa and (b) $\sigma = 200$ MPa.

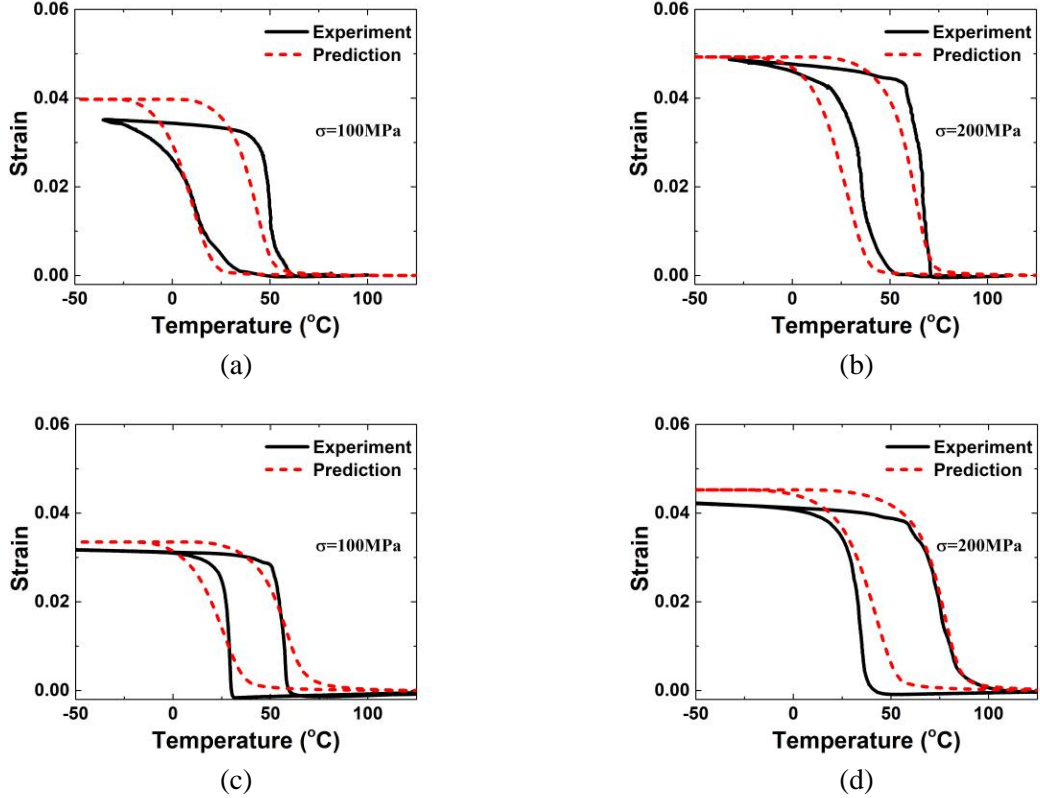

**Figure 5.** Predicted and experimental measured effective tensile strain - temperature response of the precipitated $Ni_{50.8}Ti_{49.2}$ SMAs for a thermal cycle under constant uniaxial tensile stress ($\sigma$) conditions [51]: (a) precipitates volume fraction of 1.7% and $\sigma = 100$ MPa, (b) precipitates volume fraction of 1.7% and $\sigma = 200$ MPa (c) precipitates volume fraction of 4.2% and $\sigma = 100$ MPa and (d) precipitates volume fraction of 4.2% and $\sigma = 200$ MPa.

In the same manner, in Figure 5 a comparison between the predicted response of precipitation strengthened $Ni_{50.8}Ti_{49.2}$ SMAs and the corresponding experimental data under a similar loading path is shown. Note that the used experimental results are reported in [63] while the used material parameters for the SMA matrix and preciptiates are shown in Table 1. Finally regarding the case of the preciptiation strenghened SMAs, the estimated volume fraction of the preciptiates is reported in the work of Cox *et al.* [63].

The results demostrate that the model can capture adequately the experimental response of both the homogeneous and precipitated materials. In the case of the precipitated materials, it is evident that the model successfully captures the effect of precipitation on the exhibited maximum transformation strains under different applied tensile stress conditions. In the same manner the results also show that the model predicts adequately the transformation temperatures initiation and completion points of the precipitated material under the different applied tensile stress conditions.

Finally it is important to note that in all the conducted simulations a standard RVE discretized with ~30,000 quadratic 10 node tetrahedral elements with integration at four Gauss points and hourglass control (C3D10M in [64]) is considered. This tetrahedron exhibits minimal volumetric locking during transformation and captures strain gradients in the matrix better than the standard 10-node tetrahedron due to its three extra internal degrees of freedom. The appropriateness of the used mesh density on the performed simulations was checked by discretizing one model using ~100,000 elements and comparing the overall strain–temperature response with the one obtained with the standard discretization.

## 3. BAYESIAN OPTIMAL EXPERIMENTAL DESIGN

### 3.1 Overview of the Framework

The present section describes the development of an Optimal Experimental Design (OED) framework based on the Bayesian optimization techniques. The specific objective of the framework in this study is to provide an Optimal Experiment Selection (OES) policy to guide an efficient search of precipitation strengthened NiTi SMAs with selected target properties by solving efficiently a multi-objective optimization problem. However, the general framework can be utilized for other materials discovery problems with the proper modeling tools or with the direct physical experiments. Without loss of generality, we assume that the task is to minimize the value of the '$N$' defined objective functions in a selected discrete Variables Design Space (VDS) $Z$ (i.e. this is the materials design space),

$$f_1(x) \to \min_x f_1(x),$$
$$f_2(x) \to \min_x f_2(x), \quad (5)$$
$$..., f_N(x) \to \min_x f_N(x),$$

by identifying the values of the '$M$' input variables, which are the components of the vector $x = \begin{bmatrix} X_1 & X_2 & \cdots & X_M \end{bmatrix}^T$. The aforementioned variables typically represent the degrees of freedom that can be controlled and that affect the response of the system under study. For the NiTi SMA, the considered input variables are the materials initial homogeneous Ni concentration before precipitation ($c$) and the precipitates volume fraction ($v_f$) while the objective functions are functions of the material properties of the corresponding homogenized SMA. The explicit expression of the objective functions is determined based on the operational objective(s). For example, if an SMA is queried with a targeted $A_f = 40^oC$, the objective function would have been: $f_1(c, v_f) = |A_f - 40|$. In the current work the used objective functions are provided in subsections 4.1 and 4.2. The discrete VDS, is defined a priori to the initiation of the BOED framework and it consists of '$n_T$' possible combinations of the considered variables (i.e. in the present work $c$, $v_f$). Furthermore the upper and lower bounds of the chosen VDS with respect to all the considered variables are selected based on the prior scientific knowledge about the material system of interest and its underlying physical limits. For example, in the current SMA discovery problem, $c$ and $v_f$ can take arbitrary values. However, in the examples shown in section 4, the $c$ is defined such that it ranges from 50.2 to 51.2 % at. while the input $v_f$ is defined such that it ranges from 0 to 10%. The $c$ variable bounds are selected based on the knowledge that it is not possible to form Ni$_4$Ti$_3$ precipitates below 50.2 % at. Ni and the martensitic transformation is suppressed at Ni contents above 51.2 % at. . In the same manner, the $v_f$ range is selected because, for these given Ni contents, it is impossible to obtain more than 10% precipitates. Thus, the discrete VDS is defined by selecting a desired discretization of the continuous VDS subject to the already defined bounds.

For the sake of generality in this section all the equations are presented in a generalized form and the objective is to develop a multi-objective optimization method for BOED to efficiently approach the optimal solutions of the discrete multi-objective space $S$. The optimal solutions in an optimization problem are typically referred as Pareto optimal solutions or Pareto front or Pareto front points. The Pareto optimal solutions in a selected multi-objective space, correspond to the points of the objective space that are not dominated by any other points in the same space.

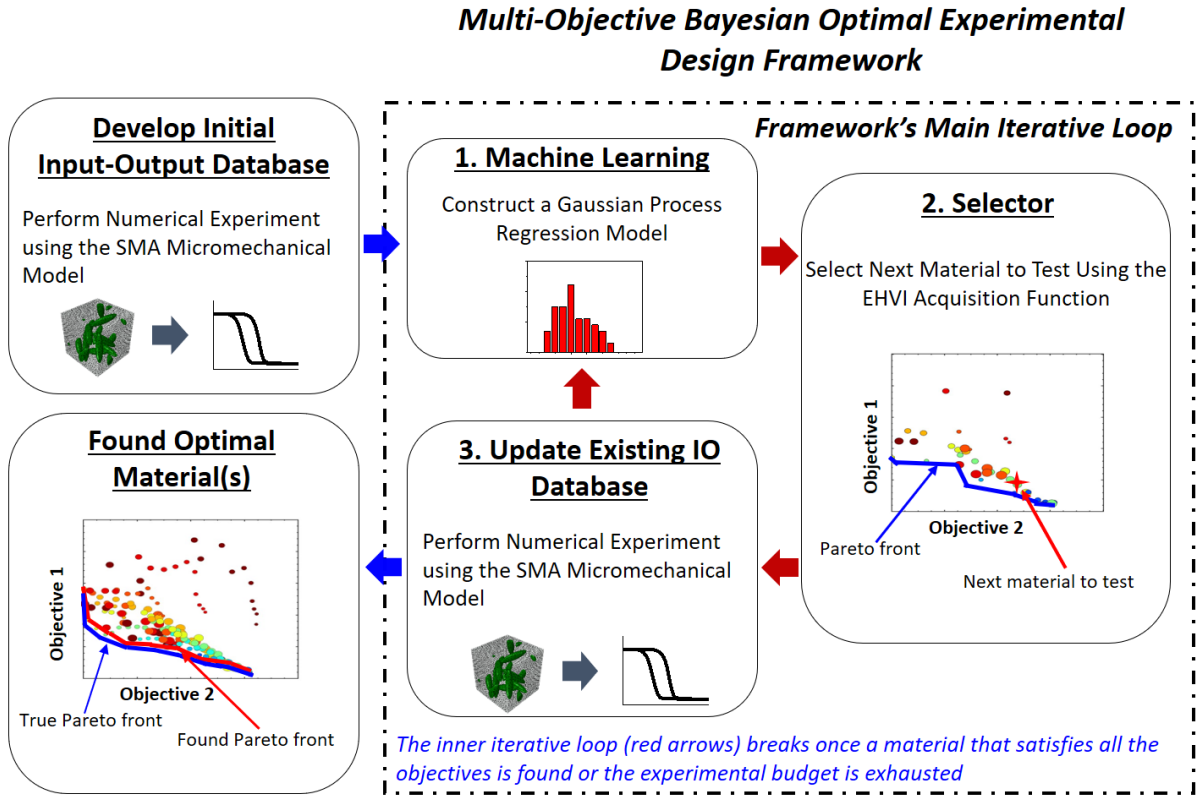

**Figure 6.** Autonomous closed-loop, multi-objective Bayesian Optimal Experimental Design framework.

In order to solve the stated problem, a BOED framework is utilized. The framework (Figure 6), operates in an iterative loop that includes three main steps: (1) the machine-learning step, (2) the selector step, and (3) the database (information)-updating step. For the BOED procedure to be initiated, '$n_I$' materials are randomly selected from the materials design space and computationally tested so that their properties are estimated and the values of the set objective functions are calculated. In the current work the SMA micromechanical model is used to perform the necessary computational experiments for each of the '$n_I$' selected materials so that their properties are calculated. Note that the '$n_I$' selected materials correspond to '$n_I$' randomly selected, without replacement, combinations of the considered input variables (i.e. in the present work $c$, $v_f$) of the system under study in the predefined materials design space of interest. These data are used in order to develop an initial database of the values of degrees of freedom and corresponding objectives functions (i.e. database of input-output variables) of the system under study.

With the completion of the initialization process, the BOED framework's main iterative loop starts with the machine-learning step, in which the collected data are used to construct a probabilistic surrogate model in order to model the input-output relationships of the system under study with uncertainty. Based on the designed acquisition function characterizing the expected utility of each candidate material with respect to this initial probabilistic surrogate model, the selector step determines the priority of the next material (i.e. any legitimate combination of input variables in the discrete VDS) to query. It is worth noting that during the selector step the determined next material to query always belongs to the unexplored materials design space due to the assumption that the results of the micromechanical model are noise-free.

For the selected next candidate material, a new set of computational experiments is conducted in order to acquire the values of the objective functions. At that point, the collected data are used in the database-updating step to update the database of the available input-output data and the Pareto front of the explored subspace of the multi-objective space $S$ is calculated and the first iteration of the BOED process is completed. If the last queried material satisfies all the targeted property requirements, the material discovery procedure can be terminated while in the case where the candidate material is not satisfactory, a new iteration of the BOED process begins. It is worth noting that in a multi-objective optimization problem it is relatively rare for a solution to satisfy all the objectives. Thus, in the current framework after the completion of each iteration of the BOED process the Pareto front of the explored subspace of the multi-objective space $S$ is calculated. The aforementioned Pareto front is not necessary the "true" Pareto front of the space $S$ and this can only be determined in cases where the ground truth is known. In the current work for the sake of simplicity the Pareto front of space $S$ will be referred as *true Pareto* front while a Pareto front calculated based on subspaces of space $S$ will be referred as Pareto front.

In each of the following iterations, during the machine-learning step the probabilistic surrogate model is updated using all the available input-output data up to that point while during the selector step the next material to query is selected out of the remaining unexplored materials design space. Finally, once the next material to test is determined, during the database-updating step, the SMA micromechanical model is used to conduct the required set of computational experiments and update the database of the input-output data. The iterations continue until the desired material is found or until a preselected number ($n_E$) of material queries is achieved guided by the BOED framework (i.e. this corresponds to equivalent iterations of the framework). The allocation among the '$n_I$' and '$n_E$' material queries is made based on the available experimental budget ($n_B = n_E + n_I$) and is selected such that the efficiency of the method is maximized. The corresponding details for the machine learning and selector steps for this work are detailed as follows.

### 3.2 Machine Learning Step

To enable multi-objective BOED, during the machine learning step, an independent GPR model [32] is chosen as the probabilistic surrogate model for the corresponding objectives. Without loss of generality, with '$n$' queried materials, the GPR can be constructed to model the materials behavior based on these data. We denote,

$$\hat{X} = \begin{bmatrix} \hat{X}_{11} & \hat{X}_{12} & \cdots & \hat{X}_{1M} \\ \hat{X}_{21} & \hat{X}_{22} & \cdots & \hat{X}_{2M} \\ \vdots & \vdots & \ddots & \vdots \\ \hat{X}_{n1} & \hat{X}_{n2} & \cdots & \hat{X}_{nM} \end{bmatrix}, \quad (6)$$

a matrix that contains the values of the input variables for the '*n*' queried materials, and

$$\hat{Y} = \begin{bmatrix} \hat{Y}_{11} & \hat{Y}_{12} & \cdots & \hat{Y}_{1N} \\ \hat{Y}_{21} & \hat{Y}_{22} & \cdots & \hat{Y}_{2N} \\ \vdots & \vdots & \ddots & \vdots \\ \hat{Y}_{n1} & \hat{Y}_{n2} & \cdots & \hat{Y}_{nN} \end{bmatrix}, \quad (7)$$

the complementary matrix that contains the values of the objective functions, modeling the materials behavior of interest. In equation (6) each row of the matrix $\hat{X}$ corresponds to a vector,

$$\hat{x}^i = \begin{bmatrix} \hat{X}_{i1} & \hat{X}_{i2} & \cdots & \hat{X}_{iM} \end{bmatrix}^T, \quad (8)$$

which contains the values of the '*M*' considered input variables. Their combination determines the candidate material used during the $i^{th}$ material query. In the same manner in equation (7) each column of the matrix $\hat{Y}$ corresponds to a vector,

$$\hat{y}_i = \begin{bmatrix} \hat{Y}_{1i} & \hat{Y}_{2i} & \cdots & \hat{Y}_{ni} \end{bmatrix}^T, \quad (9)$$

which contains the corresponding simulated values of the $i^{th}$ output variable for all the '*n*' queried materials. Finally $\hat{Y}_{ij} = f_j(\hat{x}^i)$ estimates the value of the $j^{th}$ objective function based on the $\hat{x}^i$ input vector. Note that in the current work, capital bold italic letters are used to denote matrices and lowercase bold italic letters are used to denote vectors. A vector defined by a row or a column of a matrix is denoted with the same symbol as the matrix but in a lowercase form. Superscripted or subscripted indices are used to denote vectors that correspond to a row or a column of a matrix respectively. Finally, the components of a vector or a matrix are denoted with capital italic letters using subscripted indices.

Using the observed data pairs $(\hat{X}, \hat{Y})$ of the '*n*' queried materials a GPR model is constructed to approximate the objective functions $f_j(\bullet), \forall j$ with uncertainty. To define the GPR model, a constant mean function,

$$h_c(\mathbf{a}; \boldsymbol{\theta}) = c_t, \quad (10)$$

is selected while the squared exponential kernel,

$$k(\mathbf{a}, \mathbf{a}'; \boldsymbol{\theta}) = \tau_1 exp\left[-\frac{1}{2}\frac{\|\mathbf{a}-\mathbf{a}'\|^2}{\tau_2}\right], \quad (11)$$

is chosen as the covariance function [32] in this paper, which gives a larger value with the smaller distance between two arbitrary input vectors **a** and **a**′ in equation (11). Here, $\boldsymbol{\theta} = \begin{bmatrix} \tau_1 & \tau_2 & c_t \end{bmatrix}^T$ defines the hyper-parameters of the GPR model based on which the latter equations are estimated. In the current work, the hyper-parameters of the GPR model are calculated by following the maximum likelihood (ML-II) estimation [32]. Therefore, the expression of the marginal log-likelihood [32] of the observed data $(\hat{X}, \hat{Y})$, assuming a multivariate Gaussian density function,

$$\begin{aligned} \log P_{df}(\hat{\mathbf{y}}_i / \hat{X}; \boldsymbol{\theta}^i) = \\ -\frac{1}{2}(\hat{\mathbf{y}}_i - c_t \mathbf{I}_n)^T \mathbf{K}(\hat{X}, \hat{X}; \boldsymbol{\theta}^i)^{-1}(\hat{\mathbf{y}}_i - c_t \mathbf{I}_n) \\ -\frac{1}{2}|\mathbf{K}(\hat{X}, \hat{X}; \boldsymbol{\theta}^i)| - \frac{n}{2}\log 2\pi, \end{aligned} \quad (12)$$

is maximized by a quasi-Newton method for all the considered objective functions in order to determine the corresponding hyper-parameters $\boldsymbol{\Theta} = \begin{bmatrix} \boldsymbol{\theta}_1 & \boldsymbol{\theta}_2 & \cdots & \boldsymbol{\theta}_N \end{bmatrix}$; $\boldsymbol{\theta}_1, \boldsymbol{\theta}_2$ to $\boldsymbol{\theta}_N$ correspond to hyper-parameters of the GPR model of each of the considered objective functions while the index '*i*' is used to denote variables that are calculated for, or evaluated based on, the $i^{th}$ objective function. In equation (12), $|\cdot|$ represents the determinant of a matrix; $I_n$ is a vector of length '$n$' with all the elements being 1. Furthermore $K(A, A'; \theta)$ is the covariance matrix of all possible pairs of the arbitrary matrices $A = \begin{bmatrix} \mathbf{a}_1 & \mathbf{a}_2 & \cdots & \mathbf{a}_l \end{bmatrix}$ and $A' = \begin{bmatrix} \mathbf{a}'_1 & \mathbf{a}'_2 & \cdots & \mathbf{a}'_p \end{bmatrix}$ for a given set of hyper-parameters $\theta$ and is defined as,

$$K = \begin{bmatrix} K_{11} & K_{12} & \cdots & K_{1p} \\ K_{21} & K_{22} & \cdots & K_{2p} \\ \vdots & \vdots & \ddots & \vdots \\ K_{l1} & K_{l2} & \cdots & K_{lp} \end{bmatrix}, \quad (13)$$

where the components of $K$ are calculated as $K_{ij} = k(\mathbf{a}_i, \mathbf{a}'_j; \theta)$ and $l$ and $p$ denote the number of columns of the first and second input matrices, respectively. In the same manner $h(A; \theta)$ for the arbitrary input matrix $A = \begin{bmatrix} \mathbf{a}_1 & \mathbf{a}_2 & \cdots & \mathbf{a}_l \end{bmatrix}$ given the hyper-parameters $\theta$, is defined as,

$$\mathbf{h} = \begin{bmatrix} h_1 \\ h_2 \\ \vdots \\ h_l \end{bmatrix}, \quad (14)$$

where $h_i = h_c(\mathbf{a}_i; \theta)$. Once this process is completed the values of $\Theta$ are determined and the GPR model is considered constructed as the surrogate model.

The properties of the GPR model guarantee that given the observed data pairs $(\hat{X}, \hat{Y})$ for the '$n$' queried materials, the predicted posterior distributions for the objective functions of any new experiment,

$$\tilde{\mathbf{y}}^i = \begin{bmatrix} \tilde{Y}_{i1} & \tilde{Y}_{i2} & \cdots & \tilde{Y}_{iN} \end{bmatrix}^T, \quad (15)$$

based on a new input vector,

$$\tilde{\mathbf{x}}^i = \begin{bmatrix} \tilde{X}_{i1} & \tilde{X}_{i2} & \cdots & \tilde{X}_{iM} \end{bmatrix}^T, \quad (16)$$

are also multivariate Gaussian distributions $\tilde{\mathbf{y}}^i | \tilde{\mathbf{x}}^i; \hat{X}, \hat{Y} \sim N(\mu, \Sigma)$ with mean value $\mu$ and variance $\Sigma$. In equation (15) the components $\tilde{Y}_{ij} = f_j(\tilde{\mathbf{x}}^i)$ represent the value of the $j^{\text{th}}$ objective function based on the $\tilde{\mathbf{x}}^i$ input vector and $\tilde{\mathbf{x}}^i$ is a vector that contains the values of the variables. Given the above and provided the observed data $(\hat{X}, \hat{Y})$ the values of $\mu$ and $\Sigma$ are computed for the remaining $n_R = n_T - n$ materials in the unexplored materials design space, for all the considered objective functions with the following closed-form expressions [32],

$$\begin{aligned} \mu_{ij} &= h(\tilde{\mathbf{x}}^i; \theta_j) \\ &+ K(\tilde{\mathbf{x}}^i, \hat{X}; \theta_j).. \\ &K(\hat{X}, \hat{X}; \theta_j)^{-1}... \\ &(\hat{\mathbf{y}}_j - h(\hat{X}; \theta_j)), \end{aligned} \quad (17)$$

$$\begin{aligned} \Sigma_{ij} &= K(\tilde{\mathbf{x}}^i, \tilde{\mathbf{x}}^i; \theta_j) \\ &- K(\tilde{\mathbf{x}}^i, \hat{X}; \theta_j)... \\ &K(\hat{X}, \hat{X}; \theta_j)^{-1}... \\ &K(\hat{X}, \tilde{\mathbf{x}}^i; \theta_j). \end{aligned} \quad (18)$$

In equations (17) and (18) the $i$ and $j$ indices denote the $i^{\text{th}}$ queried material and the $j^{\text{th}}$ objective function. Once the corresponding $\mu$ and $\Sigma$ values are calculated, the process continues to the selector step.

### 3.3 Selector Step

A BOED approach based on the EHVI multi-objective acquisition function is chosen to be implemented in the current framework during the selector step to guide the search of the input variables space to approach the Pareto front in the objective function space $S$. The selected acquisition function EHVI extends the idea of expected improvement for the single-objective cases, presented in [31], to the hyper-volume of the multi-objective space $S$ [15] and is defined as follows,

$$\begin{aligned} &EHVI(\tilde{\mathbf{x}}^i; \hat{X}, \hat{Y}, \Theta) = \\ &\int I_{\mathcal{H}}(\tilde{\mathbf{y}}^i | \tilde{\mathbf{x}}^i; \hat{Y}) P_{df}(\tilde{\mathbf{y}}^i | \tilde{\mathbf{x}}^i; \hat{X}, \hat{Y}, \Theta) d\tilde{\mathbf{y}}^i. \end{aligned} \quad (19)$$

In equation (19) the posterior predictive probability density function of the GPR surrogate model updated with the observed data is calculated as per,

$$P_{df}(\tilde{y}^i \mid \tilde{x}^i; \hat{X}, \hat{Y}, \Theta) = \prod_{j=1}^{N} P_{df}(\tilde{Y}_{ij} \mid \tilde{x}^i; \hat{X}, \hat{y}_j, \theta_j), \quad (20)$$

while the hyper-volume improvement function of $\hat{Y}$, $\tilde{y}^i$ at a given $\tilde{x}^i$ is calculated as per,

$$I_{\mathcal{H}}(\tilde{y}^i \mid \tilde{x}^i; \hat{Y}) = \mathcal{H}(\hat{Y} \cup \tilde{y}^i) - \mathcal{H}(\hat{Y}), \quad (21)$$

where $\cup$ denotes matrix concatenation. In equation (21) the hyper-volume dominated by the approximation set $\hat{Y}$ and $\hat{Y} \cup \tilde{y}^i$ are defined based on the expression,

$$\mathcal{H}(B) = \mathrm{Vol}\left(\{\alpha \in \mathbb{R}^N \mid \alpha \prec r \text{ and } \exists \beta \in B : \beta \prec \alpha\}\right), \quad (22)$$

where $B$ denotes the arbitrary input matrix, $\beta \in B$ indicates that $\beta$ is a row vector of $B$ and $r$ is a point dominated by all the points in $S$, called the reference point. In other words assuming that the observed data sequence is $\hat{Y}$, then $\mathcal{H}(\hat{Y})$ and $\mathcal{H}(\hat{Y} \cup \tilde{y}^i)$ are evaluated based on equation (22) and denote the hyper-volume dominated by the Pareto front of the current set $\hat{Y}$ and $\hat{Y} \cup \tilde{y}^i$ respectively.

In the current framework once the surrogate model is constructed based on the observed '$n$' data, the EHVI is evaluated for the '$n_R$' remaining combinations of the considered variables in the predefined VDS. The new selected point is defined as $x^* = \arg\max_{\tilde{x}^i \in \tilde{X}} EHVI(\tilde{x}^i; \hat{X}, \hat{Y}, \Theta)$ where $\tilde{X}$ denotes a matrix that contains the '$n_R$' remaining combinations of the considered variables,

$$\tilde{X} = \begin{bmatrix} \tilde{X}_{11} & \tilde{X}_{12} & \cdots & \tilde{X}_{1M} \\ \tilde{X}_{21} & \tilde{X}_{22} & \cdots & \tilde{X}_{2M} \\ \vdots & \vdots & \ddots & \vdots \\ \tilde{X}_{n_R 1} & \tilde{X}_{n_R 2} & \cdots & \tilde{X}_{n_R M} \end{bmatrix}. \quad (23)$$

Finally, in the current work, a Hypervolume based Pure Exploitation (HPE) acquisition function,

$$HPE(\tilde{x}^i; \hat{X}, \hat{Y}, \Theta) = \mathcal{H}(E(\tilde{x}^i \mid \tilde{y}^i; \hat{X}, \hat{Y}, \Theta)), \quad (24)$$

is also considered where $E(\tilde{x}^i \mid \tilde{y}^i; \hat{X}, \hat{Y}, \Theta)$ is the posterior mean of $\tilde{y}^i$. This acquisition function, calculates the hypervolume dominated by the posterior mean of $y^i$, without considering the posterior variance as the EHVI acquisition function does. This acquisition function will be also used within the BOED framework for comparison purposes to provide a PEES policy to guide a search of precipitation strengthened NiTi SMAs with selected target properties. As it will be demonstrated, this acquisition function biases the search towards the optimality of the desired performance under the current surrogate model.

## 4. RESULTS AND PERFORMANCE ANALYSIS

In this section, the capabilities of the developed BOED framework are demonstrated by solving two distinct materials discovery problems. The first problem deals with the query of a precipitation strengthened NiTi SMA that satisfies two objectives, while the second problem deals with the query of a precipitation strengthened NiTi SMA that its properties satisfy three objectives. The explicit expressions of the adopted objectives for the two problems are defined in the subsections 4.1 and 4.2, respectively and were selected for specific aerospace applications in order to use NiTi SMAs as solid-state actuators. In both problems an SMA material is queried in the domain of the materials space defined by the used variables $c$ and $v_f$ as follows: $c$ ranges from 50.2 to 51.2 % at. and $v_f$ ranges from 0 to 10% (i.e. these are the selected variables bounds which define the continuous VDS).

## 4.1 Materials Discovery – 2 Objectives Problem

In this example the developed BOED framework is used to discover a precipitation strengthened NiTi SMA with (objective 1) an austenitic finish temperature $A_f = 30^oC$ and a (objective 2) specific thermal hysteresis that is defined by the difference of austenitic finish temperature and martensitic start temperature, $A_f - M_s = 40^oC$. The stated problem is solved for two case studies, where the selected continuous VDS is discretized with a coarse and a dense mesh respectively.

### 4.1.1 Case Study 1 – Coarse Discretization of the Continuous Variables Design Space

In this case study, a coarse discretization of the continuous VDS is chosen such that the discrete VDS it consists of $n_T = 231$ combinations of the considered variables $c$ and $v_f$ and is defined as follows: $c$ ranges from 50.2 to 51.2 % at. with the increment of 0.1 % at. and $v_f$ ranges from 0 to 10% with increment 0.5%. The selected discretization is chosen based on the experimentally observed sensitivity of the materials response to the Ni concentration and microstructural heterogeneities in NiTi SMAs. This case is utilized to demonstrate the capabilities of the developed method, in an example case where the true Pareto front of the objective space is known by carrying out first an HT analysis. Thus, the results of the HT analysis facilitate the quantification of the efficiency of the BOED framework in detecting the true Pareto front. Furthermore, the performance of the OES policy in detecting the true Pareto front within a predefined experimental budget is also quantified and compared with the performance of the PRES policy.

To demonstrate the benefits of the employed BOED framework, an HT approach is carried out first where all the materials belonging to the materials design space are tested and their properties are analyzed so that the true Pareto front of the objective space is calculated. Figure 7 shows the corresponding results of the HT approach and it demonstrates the values of the materials properties of interest, transformation temperatures in this case, in the different regions of the whole design variables space. Note that the white regions in Figure 7 correspond to values of the variables space which correspond to nonphysical materials. That is because using these values, in equation (2) the calculated value of $\bar{c}_A$ corresponds to a Ni concentration in the SMA matrix, after precipitation, which is higher than the initial Ni concentration of the homogenous material; a condition which is not physically possible and therefore a corresponding material with these $c$ and $v_f$ values cannot exist. Therefore, it is calculated that 54 combinations of the considered variables correspond to nonphysical materials and therefore the VDS reduces to $n_T = 174$ points. The results shown in Figure 7 are used to calculate the values of the objective functions as a function of the design variables and are illustrated in Figure 8. It is interesting to note in the figure that there does not exist any region in the VDS that simultaneously optimizes both of the objective functions. This result is anticipated, since the problem here is a nontrivial multi-objective (2-objective in this case) optimization problem and there is no single optimal solution that simultaneously satisfies both objectives. In fact, there exists a set of Pareto optimal solutions which represent the true Pareto front of the selected objective space which in other words these points represent the optimal materials based on the selected objectives. In order to visualize the true Pareto front of the objective space using the response surfaces illustrated in Figure 8, the values of the objectives for the VDS are shown in Figure 9, with the corresponding true Pareto front that consists of 21 data points in this case. It is worth pointing out that the identification of the true Pareto front of the objective space would have been impossible without the prior knowledge of the response surface. Hence this highlights one of the contributions of the present work which is the demonstration of a framework that can approach the Pareto front of a problem assuming that (1) there is no prior knowledge of the response surface and (2) each query of the problem is expensive and thus a thorough exploration of the design space is not feasible.

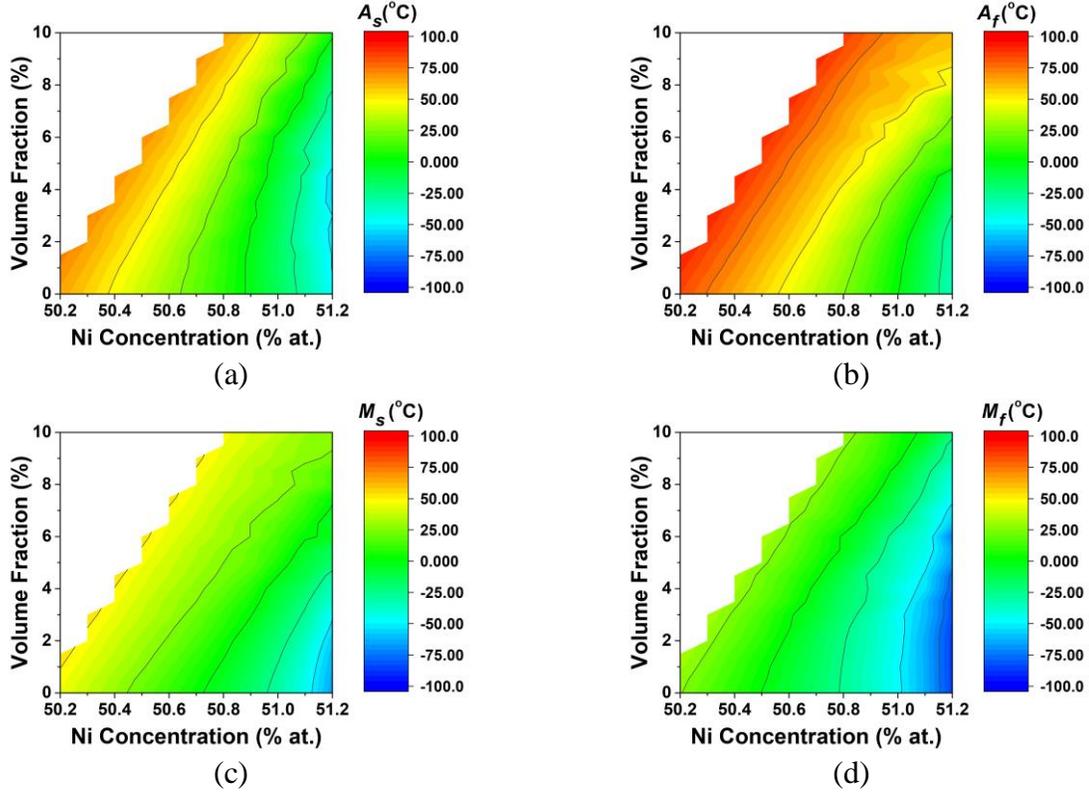

**Figure 7.** Effects of the precipitate volume fraction and initial Ni concentration on the transformation temperatures of NiTi SMAs: (a) $A_s$, (b) $A_f$, (c) $M_s$ and (d) $M_f$.

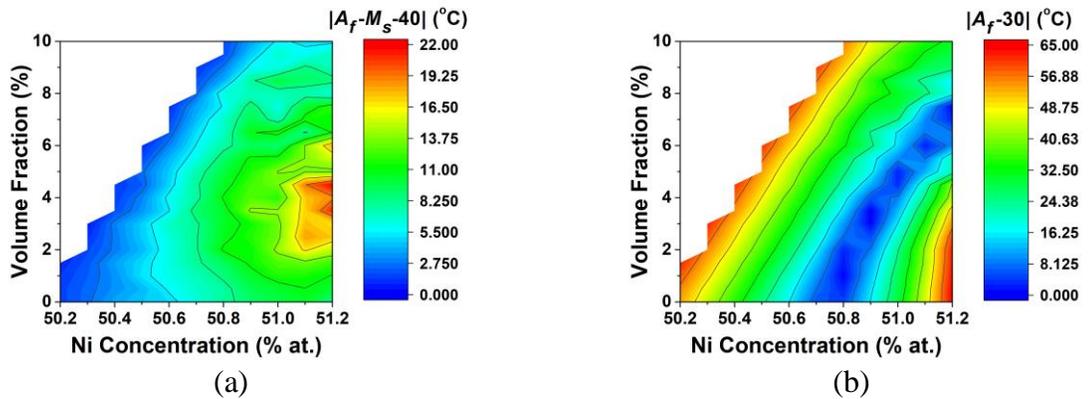

**Figure 8.** Effects of the precipitate volume fractions and initial Ni concentration on the selected objectives: (a) objective 1: $|A_f - M_s - 40| = 0$ and (b) objective 2: $|A_f - 30| = 0$.

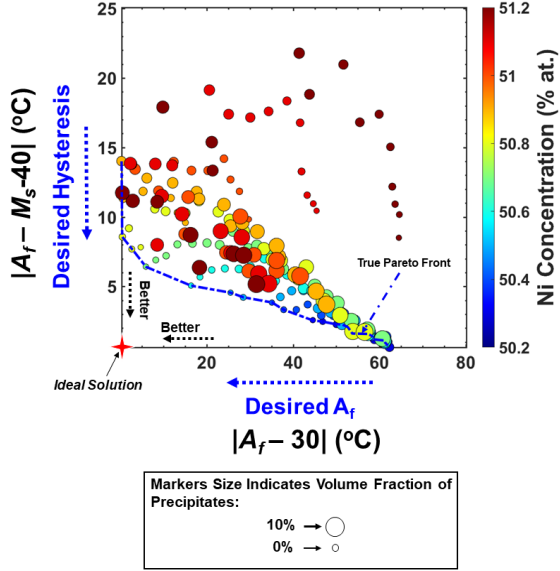

**Figure 9.** Calculated objective space and true Pareto front using the HT approach for the VDS.

To analyze the efficiency of the developed BOED framework (see Appendix B for implementation details), the problem solved exhaustively using the HT approach, is solved again with the BOED framework. The efficiency of the framework is then assessed by comparing the required number of material queries that have to be conducted using the BOED framework with the number of material queries that have to be conducted using the HT approach so that the true Pareto front can be identified. Note that in the current work for practical reasons we aim to detect the 95% of the true Pareto front points. Furthermore, the performance of the BOED framework in detecting the true Pareto front within a predefined experimental budget is evaluated under different allocations of '$n_I$' and '$n_E$' material queries. To this end, an experimental budget is assumed of $n_B = n_I + n_E = 81$ material queries and five cases of different numbers of '$n_I$' material queries are considered: $n_I = 1, 11, 21, 41$ and $61$.

For each of the aforementioned cases, the remaining experimental budget is spent on optimally guided queries of materials ($n_E = 80, 70, 60, 40$ and $20$), while in addition, a case where a PRES policy is also considered ($n_I = 81, n_E = 0$) for comparison purposes. For each of the aforementioned six cases ($n_I = 1, 11, 21, 41, 61$, and $81$), the number of true Pareto front points found after the completion of each material query ($n$) is recorded. Furthermore, to study the effect of the randomly queried materials on the performance of the BOED framework in detecting the true Pareto front points by the completion of the experimental budget, each case is evaluated 1000 times.

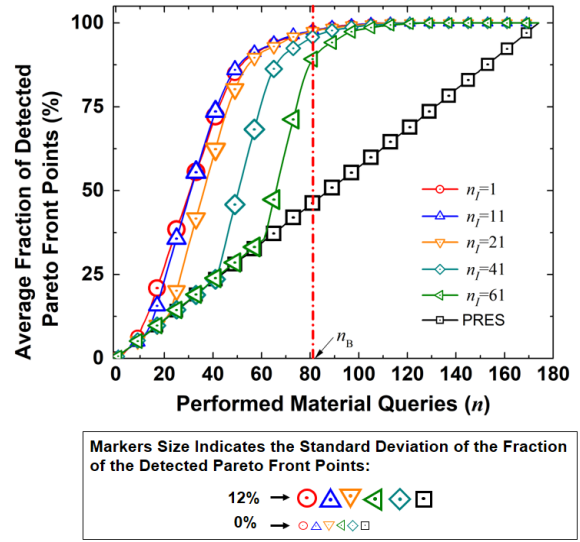

**Figure 10.** Average fraction and standard deviation of detected true Pareto front points using the BOED framework and the PRES policy during the $n^{th}$ material query. The results are shown for different configurations of the BOED framework where different numbers of '$n_I$' material queries are used. All the considered cases are evaluated 1000 times.

Therefore the average number of detected true Pareto front points and their Standard

Deviation (SD), after each material query ($n$), for each of the aforementioned cases until the completion of the experimental budget are shown in Figure 10 and Figure 11. For the sake of completeness, the figures also show the average number of detected true Pareto front points and their SD beyond the completion of the experimental budget $n_B = 81$.

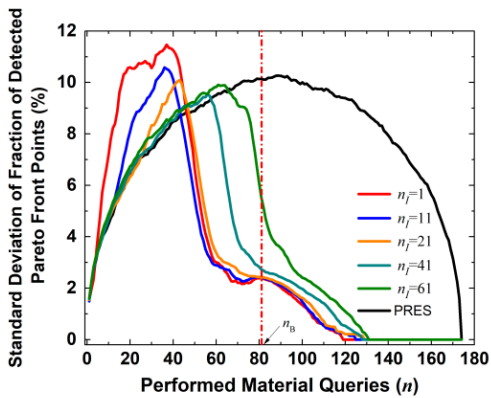

**Figure 11.** Standard Deviation of detected true Pareto front points using the BOED framework and the PRES policy during the $n^{th}$ performed set of computational experiments. The results are shown for different configurations of the BOED framework where different numbers of '$n_I$' material queries are used. All the considered cases are evaluated 1000 times.

The results indicate that in all the cases that the BOED framework is utilized, including the case of $n_I = 61$ materials queries, where a large number of randomly queried materials is performed, it finds on average, more than 90% of the true Pareto front points within the experimental budget. The results also suggest that the overall performance of the BOED method in finding the majority of the true Pareto front points, within the experimental budget is stronger, when the given experimental budget is allocated towards more optimally queried materials rather than to randomly queried materials. That is because in the cases where $n_I = 1$, 11, 21 and 41 materials queries are performed, on average, 98%, 97%, 97% and 94% of the true Pareto front points are detected within the experimental budget respectively with corresponding SD values, 2.3%, 2.4%, 2.4% and 2.7% respectively while in the case where $n_I = 61$ materials queries are performed the method detects on average 90% of the true Pareto front points with corresponding SD value 5.5%. Note that a higher value of average fraction of detected Pareto front in conjunction with a lower value of SD indicates a better overall performance. Finally, the same figure shows that the PRES policy ($n_I = 81$) only detects lower than half of the true Pareto front points within the experimental budget with corresponding SD value 10% and it exhibits an inferior performance in finding the true Pareto front points within the experimental budget than all the cases where the BOED framework is adopted.

The present results demonstrate that, for the specific materials discovery problem at hand, by employing the BOED framework and allocating a reasonable fraction of the experimental budget on tests on material suggested by the developed method, the majority of the true Pareto front points can be detected. Additionally, the results also show that the BOED framework with different setups (i.e. with different values of '$n_I$') can always perform better than the PRES policy.

Furthermore, the present BOED framework, as compared against the HT approach, exhibits, on average, an improvement on the efficiency of nearly 60% in the case where $n_I = 1$ material queries are performed. That is because in that case it requires the conduction of nearly 60% less material queries in comparison to the HT approach (i.e. 68 sets of experiments in comparison to 166) in order to find, on average, the 95% of the true Pareto front points. Furthermore, it is interesting to note that the efficiency of the BOED method reduces as the number of the randomly queried materials increases. The results shown in Figure 10 indicate that for the BOED method to detect on average the 95% of the true Pareto front points, for the cases of the $n_I = 11, 21, 41$ and 61 randomly queried materials, an experimental budget of $n_B = $ 69, 71, 79 and 91 material queries on optimally selected materials, respectively, would have been required. Such a trend indicates the reduction of the efficiency of the BOED method in comparison to the case where $n_I = 1$ materials are randomly selected and tested.

Finally the results shown in Figure 10 and Figure 11 also indicate that for a selected $n_I$ value, as the experimental budget increases further than the selected $n_B = 81$, the average performance of the BOED framework in finding the majority of the true Pareto front points improves. Additionally in the same manner, the effect of the randomly queried materials on the consistency of the predictions of the method becomes less prominent (lower SD values) regardless the variability of selected materials. In the same manner as the experimental budget reduces the average performance of the method is reducing while its results are becoming more sensitive on the randomly selected materials. However it is interesting to note that even in the cases of low experimental budgets (i.e. $n_B \leq 40$), the results shown in Figure 10 and Figure 11 indicate that the BOED framework utilized while using the lowest number of randomly queried materials ($n_I = 1$) offers the best overall performance.

The above results seem to be counter intuitive since common sense would lead one to think that a constructed GPR model based on the results of a vast number of randomly queried materials will perform better and it will minimize the number of the optimally queried materials in order to detect the true Pareto optimal solution. Indeed, our results (Figure 10) suggest that the efficiency of the method at detecting, consistently, the majority of the true Pareto front points, within an experimental budget, is maximized when the BOED framework is started with as few as 1 initial test on a randomly selected material. This is because, in such a case, each of the subsequent optimally queried materials provide the most information about the BOED, in regions of the materials design space that are most likely to yield a (multi-objective) desired response. This last result puts into question the utility of high-throughput approaches, particularly when accounting for limitations in available resources to carry out the physical or computational experimental tests.

### 4.1.2 Case Study 2 - Dense Discretization of the Continuous Variables Design Space

In this case study the selected VDS is refined in order to have $n_T = 21021$ combinations of the considered variables $c$ and $v_f$. Thus the VDS is defined as follows: $c$ ranges from 50.2 to 51.2 % at. with increment 0.001 % at. and $v_f$ ranges from 0 to 10% with increment 0.05%. At this point it should be noted that the true Pareto front of case studies 1 and 2 are expected to be similar. That is because the selected discretization of the selected continuous VDS in case study 1 is already adequate enough in order to reflect the response of the material in the selected continuous space. However the main objective of this case study is to compare the utility of the queried materials by the OES, the PRES and the PEES policies within a predefined experimental budget in a problem with a relatively large variables space.

In the current materials discovery problem, an experimental budget of $n_B = 20$ material queries is selected and the aforementioned strategies are used to perform a "blind" search for an SMA with the desired properties. In this example, the experimental budget of the OES and PRES policies is allocated to $n_I = 1$ randomly queried material and to $n_E = 19$ optimally queried materials. It is worth noting that from the defined variables space, it is calculated that 5081 combinations of the considered variables correspond to nonphysical materials and therefore the variables space reduces to 15940 points.

The results of the employed strategies are demonstrated in Figure 12 where the calculated objective space is depicted.

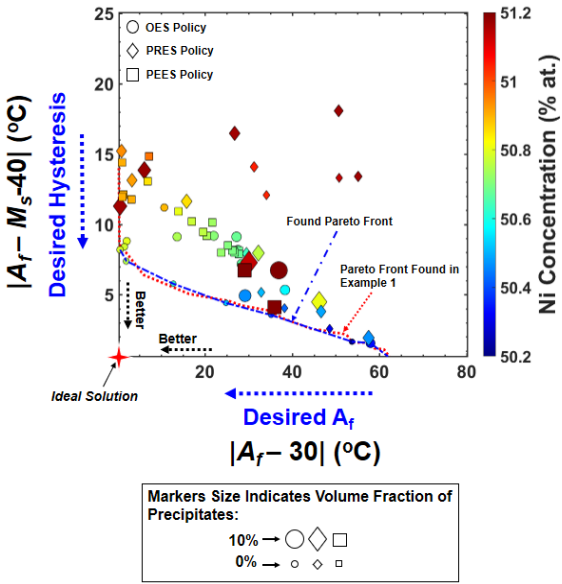

**Figure 12**. Calculated objective space and Pareto front using the OES ($n_I = 1, n_E = 19$), PRES ($n_I = 20, n_E = 0$) and PEES ($n_I = 1$, $n_E = 19$) policies under the $n_B = 20$ experimental budget.

The results indicate that the OES policy, even under this limited experimental budget, queries materials that belong to the region of the objective space which approaches the true Pareto front. This is clearly noticeable by comparing the Pareto front calculated based on the results of the OES (blue dash line) with the true Pareto front found during the case study 1 (red dot line).

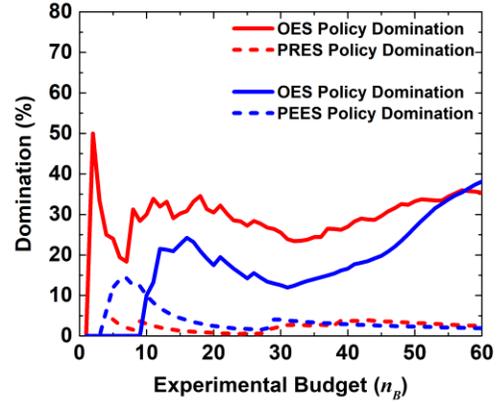

**Figure 13.** Comparison of the utility of the queried materials by the OES, PRES and PEES policies as function of the experimental budget for the 2-objectives materials discovery problem. The comparison is performed for a dense discretization of a continues VDS with predefined variables bounds: (a) The red solid curve indicates the average percentage of domination of each material queried by the OES policy over the materials queried by the PRES policy, (b) the red dash curve indicates the average percentage of domination of each material queried by the PRES policy over the materials queried by the OES policy, (c) the blue solid curve indicates the average percentage of domination of each material queried by the OES policy over the materials queried by the PEES policy and (d) the blue dash curve the average percentage of domination of each material queried by the PEES policy over the materials queried by the OES policy.

The results also show that the materials queried by the PRES policy are randomly dispersed in the objective space as is expected while the materials queried by the PEES policy are concentrated in a specific region of the objective space which consists of materials with similar volume fraction values. The behavior of the latter policy is anticipated due to its pure exploitation nature. To quantitatively compare the utility of the queried materials by the three deployed policies the results of Figure 12 are further analyzed. It is estimated that at the completion of the selected experimental budget, each material queried by the OES policy dominates on average 30% of the materials queried by the PRES policy, while the materials queried by the PRES policy dominate 1% of the materials found using the OES policy. Furthermore, the results also suggest that each material queried by the OES policy, does not dominate the 69% of the queried materials by the PRES policy and vice versa. In the same manner the results of Figure 12 also show that each material queried by the OES on average dominate 17% of the materials queried by the PEES policy, while the materials queried following the PEES policy dominate 2.5% of the materials queried using the OES policy. The aforementioned numbers indicate that the OES on average queries materials with better utility in comparison to the other two policies, while the PRES policy exhibits the worst performance. It is interesting to note that the same trends of performance are maintained through the equivalent comparisons conducted for various experimental budgets that are depicted in Figure 13. Finally, Figure 14 demonstrates that if the OES policy is employed to query a material in a discrete VDS with defined variables bounds, its relative performance in comparison to the PRES policy is more definitive as the discretization of the VDS is further refined, as the gap between PRES and OES policies for the case of the dense discretized VDS (red lines) is much bigger than that in the case of the coarse discretized VDS (blue lines).

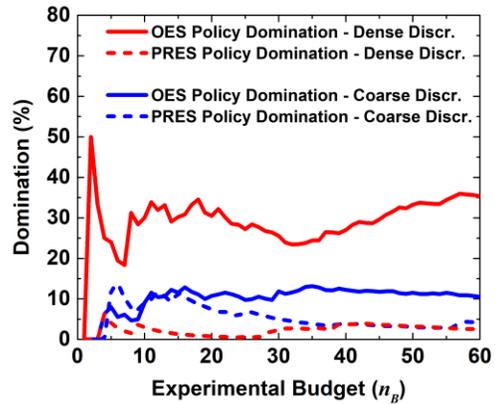

**Figure 14.** Comparison of the utility of the queried materials by the OES and PRES policies as function of the experimental budget for the 2-objectives materials discovery problem. The comparison is performed for a two discretizations of a continues VDS with predefined variables bounds: (a) The red solid curve indicates the average percentage of domination of each material queried by the OES policy over the materials queried by the PRES policy under a dense discretization of the continues VDS, (b) the red dash curve indicates the average percentage of domination of each material queried by the PRES policy over the materials queried by the OES policy under a dense discretization of the continues VDS, (c) the blue solid curve indicates the average percentage of domination of each material queried by the OES policy over the materials queried by the PEES policy under a coarse discretization of the continues VDS and (d) the blue dash curve the average percentage of domination of each material queried by the PEES policy over the materials queried by the OES policy under a coarse discretization of the continues VDS.

## 4.2 Materials Discovery – 3 Objectives Problem

In this example, the developed BOED framework is used to discover precipitated SMAs with (objective 1) an austenitic finish temperature $A_f = 30°C$, (objective 2) a specific thermal hysteresis that is defined by the difference of austenitic finish temperature and martensitic start temperature, $A_f - M_s = 40°C$ and (objective 3) the maximum transformation strain ($H_{sat}$) that the material can exhibit being maximized. For the present problem, the VDS is chosen such that it consists of $n_T = 21021$ combinations of the considered variables $c$ and $v_f$ and is defined as follows: $c$ ranges from 50.2 to 51.2 % at. with increment 0.001 % at. and $v_f$ ranges from 0 to 10% with increment 0.05%.

In this high-dimensional problem with an even more complicated response surface in comparison to the first problem, a "blind" search of the SMA with desired properties is performed. Furthermore, in addition to the OES policy, the PRES and the PEES policies are also used and the utility of the queried materials by each of the aforementioned policies within a predefined experimental budget is compared. Thus an experimental budget is assumed of $n_B = 50$ material queries and for the case of the OES and PEES policies the experimental budget is allocated to $n_I = 1$ randomly queried material and to $n_E = 49$ optimally queried materials. By analyzing the results at the completion of the selected experimental budget, each material queried by the OES policy dominates on average 39% of the materials queried by the PRES policy, while the materials queried by the PRES policy *dominate only 2%* of the materials found using the OES policy. Furthermore, each material queried by the OES policy dominates on average 47% of the materials queried by the PEES policy, while the materials queried following the PEES policy *dominate 1%* of the materials queried using the OES policy. Figure 15 shows the aforementioned results, while also shows the results for various experimental budgets. The findings indicate that the materials queried following the OES policy, when a reasonable experimental budget is used, outperform the materials queried by the other two policies. Finally, in Table 2 the queried materials that belong to the Pareto front found after the completion of the experimental budget are listed.

**Table 2.** Queried materials belonging to the Pareto Front for the 3-objectives materials discovery problem

| Ni (% at.) | $v_f$ (%) | $A_s$ (°C) | $A_f$ (°C) | $M_s$ (°C) | $M_f$ (°C) | $H_{sat}$ (%) |
|---|---|---|---|---|---|---|
| 50.2 | 0.0 | 66 | 84 | 45 | 26 | 5.1 |
| 50.331 | 3.5 | 72 | 92 | 53 | 34 | 4.7 |
| 50.395 | 0.0 | 48 | 66 | 29 | 9 | 5.1 |
| 50.53 | 6.5 | 70 | 92 | 53 | 33 | 4.4 |
| 50.715 | 0.0 | 18 | 34 | 1 | -18 | 5.1 |
| 50.761 | 10.0 | 70 | 91 | 53 | 35 | 4.0 |

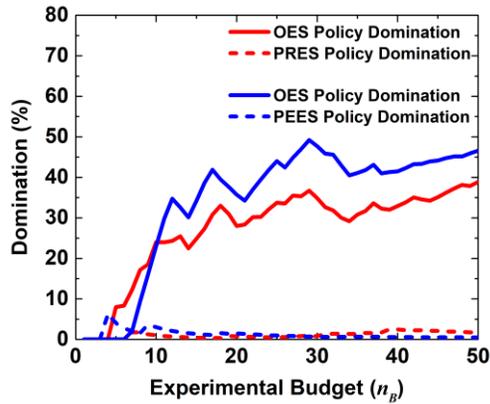

**Figure 15.** Comparison of the utility of the queried materials by the OES, PRES and PEES policies as function of the experimental budget for the 3-objectives materials discovery problem. The comparison is performed for a dense discretization of a continues VDS with predefined variables bounds: (a) The red solid curve indicates the average percentage of domination of each material queried by the OES policy over the materials queried by the PRES policy, (b) the red dash curve indicates the average percentage of domination of each material queried by the PRES policy over the materials queried by the OES policy, (c) the blue solid curve indicates the average percentage of domination of each material queried by the OES policy over the materials queried by the PEES policy and (d) the blue dash curve the average percentage of domination of each material queried by the PEES policy over the materials queried by the OES policy.

## 5. CONCLUSIONS

In the present work, a closed-loop BOED framework has been developed and implemented to guide an efficient search of precipitated SMAs with targeted properties that satisfy more than one objectives. An independent GPR model has been used as the probabilistic surrogate model during the machine learning step of the framework, to emulate the response (and its uncertainty) of the computational experiments. During the selector step a scalar EHVI multi-objective acquisition function has been used to select, at every iteration of the process, the best material to query, given the data acquired so far as well as the objective(s) of the optimization.

Finally, to develop/update the database of the input-output relationships connecting the behavior of the SMA candidate to its microstructural characteristics ($c$, $v_f$) a high fidelity SMA micromechanical model has been integrated into the framework. The ability of the SMA model to predict the response of the precipitated SMAs has been demonstrated through comparison with experimental results.

The BOED framework have been utilized to carry out an efficient search of precipitation hardened SMA materials with desired properties. Two multi-objective materials discovery problems have been considered and the found optimal materials has been reported. Furthermore, the results of the BOED framework have demonstrated that the method could efficiently approach the true Pareto front of the objective space of the approached materials discovery problems successfully. The results also indicate that the efficiency and consistency of the BOED framework in approaching the Pareto front points are maximized when the given experimental budget is allocated preferentially to the optimal material queries rather than to the initial random material queries which serve for the initial construction of the surrogate model. Furthermore, it has been demonstrated that the queried materials by the BOED framework within a predefined experimental budget exhibit a higher utility than the corresponding queried materials following the PRES and PEES policies. Finally, the decision to use the SMA micromechanical model in the BOED framework has significantly improved the overall performance of the framework, because it has enabled the implementation of a fully autonomous closed-loop approach. The in-house code developed in the current work related to the BOED framework along with the user's manual and the generated data are shared through a GitHub repository [65].

In conclusion, the present results: (a) demonstrate the effectiveness of a closed-loop BOED framework to optimize multiple materials performance metrics simultaneously, (b) highlight the efficiency of the utilized methods in comparison to the conventional high throughput approaches, (c) demonstrate the capability of the framework to inverse a high-fidelity model with no closed- form solution and (d) identify precipitation hardened SMAs with desired properties suitable for aerospace related actuation applications.

Future work will focus on further experimental validation of the developed framework. Additional work will be focused on the fusion of computational and experimental data in the developed BOED process. Thus, physical experiments will be conducted during selected iterations of the framework in order to quantify to improve the predictive capabilities of the used SMA micromechanical model. The quantify uncertainty of the predictions of the latter model, in turn will be taken into consideration in the BOED framework to improve the accuracy of the material queries. Additional efforts will be directed on the application of the developed framework on the discovery of precipitated polycrystalline NiTiHf SMA ternary systems. This task will include the development of SMA micromechanical models for the NiTiHf material system as well as the use of alternative acquisition functions within the BOED framework to further improve the efficiency of the method. Further development of this framework will include the incorporation of processing-microstructure linkages within the same unified BOED approach.

# ACKNOWLEDGEMENTS

The material is based upon work supported by the NSF Grant 'DMREF: Accelerating the Development of High Temperature Shape Memory Alloys' under the Award Number: 1534534 and by the NSF Grant 'CAREER: Knowledge-driven Analytics, Model Uncertainty, and Experiment Design' under the Award Number: 1553281. Evaluations of the thermo-mechanical behavior of the precipitated polycrystalline NiTi SMAs were carried out in the Texas A&M Supercomputing Facility.

# APPENDIX A

*Calculation of Material Parameters of a Homogenized SMA*

In this section, the procedure followed to determine the material parameteres of a homogenized SMA using the predicted, by the SMA micromechanical model, response of a precipitated polycrystalline SMA is described. The process follows the SMA material characterization procedure desribed in [59], [61] which was originally developed to characterize the response of SMAs through physical experiments and to calculate the material parameters of the polycrystalline SMA constitutive model developed by Lagoudas *et al.* [59]. In the current work the same aproach is adopted while the physical experiments are replaced by computational experiments conducted by the SMA micromechanical model.

To determine the material parameteres of a homogenized SMA, based on the response of a precipitated polycrystalline SMA of known precipitate volume fraction and initial Ni concenetration, the SMA constitutive model developed by Lagoudas *et al.* [59] is used and it is calibrated following the material characterization procedure described in [59], [61]. To this end, the developed micromechanical model is used and is subjected into: (a) thermal cycling loading paths under four constant uniaxial tensile stress conditions: 100, 200, 300 and 400 MPa and (b) a mechanical loading–unloading path under tensile stress conditions, sufficient to fully transform the material, at constant nominal temperature higher than the austenitic finish temperature. The first loading path type is used to get the isobaric response of the material, under diffrent constant uniaxial loads, in order to construct the phase diagram (Figure 16a).

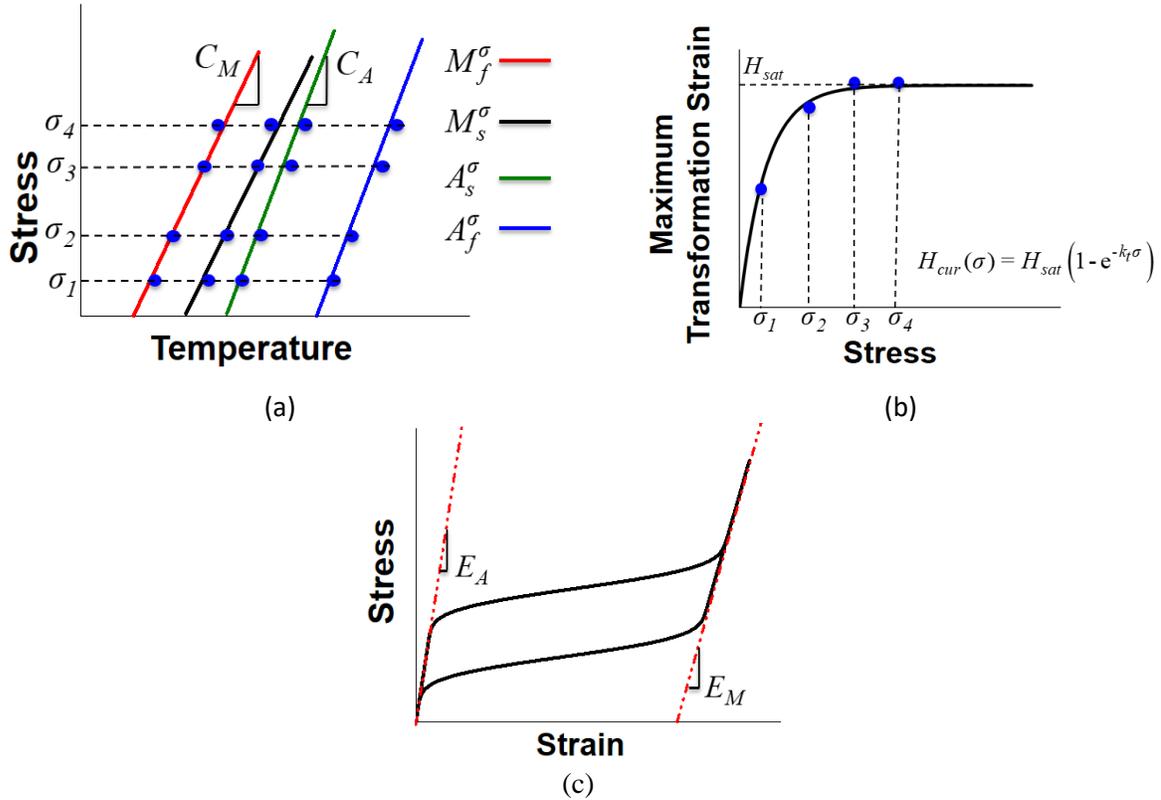

**Figure 16.** Material characterization process. (a) Calculation of transformation temperatures and stress influence coefficients by using materials phase diagram; the phase diagram is constructed by using the isobaric response of the material under different uniaxial tensile loads. (b) Calculation of the exhibited maximum transformation strain as a function of the applied constant uniaxial tensile stress (c) Calculation of elastic modulus of austenite and martensite from an SMA pseudoelastic response.

The SMA phase diagram is used to calculate the transformation temperatures at zero stress $(M_s, M_f, A_s, A_f)$ and the stress influence coefficients $(C_A, C_M)$ of the homogenized material. Furthermore the same data are used to acquire the parameters $H_{sat}$ and $k_t$ of the function $H_{cur}(\sigma) = H_{sat}\left(1-e^{-k_t\sigma}\right)$ which relate the maximum transformation strain ($H_{cur}$) at current stress with the applied constant stresses ($\sigma$) (Figure 16b). The elastic modulus of austenitic ($E_A$) and martensitic phases ($E_M$) of the precipitated material are subsequently derived by using its pseudoelastic response (Figure 16c) acquired by the second loading path. That is achieved by considering a material which is initially at the austenitic phase and by loading it uniaxially at a nominally constant temperature higher than $A_f$ until it transforms to the martensitic phase. Subsequent unloading induces reverse phase transformation back to the austenitic phase. The slopes of the two elastic regions in the dervied plot depict the elastic modulus of austenitic and martensitic phases respectivly. Once the process is completed the material properties $\left(M_s, M_f, A_s, A_f, C_A, C_M, E_A, E_M, H_{\text{sat}}, k_t\right)$ of the material under study are calculated. This process is repeated for three different dispersions of 15 precipitates and the average material properties are considered representative of the material-at-large [52]. Details about the utilization of the parallel computing capabilities of the ADA high performance research computing cluster at Texas A&M University in order to perform in a computationally efficiently manner the needed computational experiments to calculate the material parameteres for a homogenized SMA model are described in Appendix C.

## APPENDIX B

*Implementation of the BOED framework*

An in-house implementation of the BOED framework has been developed by the authors. The flowchart of the algorithm along with the inputs and outputs of the process is shown in Figure 17. Note that for the sake of simplicity the flowchart shown in Figure 17 depicts the algorithm followed for the solution of the materials discovery problem discussed in Section 4.1 of the manuscript. A MATLAB [66] script has been used to drive the process while the machine learning and selector parts of the algorithm are also developed in the same programing platform. A Python script compatible with the ABAQUS FE software [64] scripting interface has been developed to generate the FE based SMA micromechanical models and the corresponding ABAQUS input files based on the values of the provided $c$ and $v_f$ variables and the supplementary parameters required by the SMA micromechanical model. Thus, the ABAQUS FE software has been used to run the needed FE based micromechanical simulations where a User Material subroutine (UMAT) has been developed in FORTRAN programing language to model the behavior of the SMA matrix based on the polycrystalline SMA constitutive model of Lagoudas et al. [59]. Finally, provided the results of the SMA micromechanical model a homogenized SMA is determined using a MATLAB script following the procedure described in Appendix A.

In order to perform the BOED using the developed in-house code, the following inputs should be provided to the code. 1) The predefined design variables space $c$ and $v_f$, 2) the objective functions, which for the presented case are the $f_1(c, v_f) = |A_f - M_s - 40|$ and $f_2(c, v_f) = |A_f - 30|$, 3) the values of the parameters $n_I$ and $n_E$, and 4) the parameters required by the SMA micromechanical model. During the initiation of the process, an initial database of the input-output variables of the problem is developed. In this case the input variables are the $c$ and $v_f$ and the output variables are the $f_1(c, v_f)$ and $f_2(c, v_f)$. To this end the red marked iterative process shown in the flowchart in Figure 17 starts by selecting randomly without replacement a pair of input variables from the predefined VDS. During the following step the calculation of operational objectives is performed following the steps shown in Figure 18. Thus, the ABAQUS python script is executed so that the ABAQUS input file is generated and the latter is used to perform five computational experiments under different thermo-mechanical loading conditions following the procedure described in Appendix A of the manuscript. The acquired isobaric and pseudoelastic responses of the material are utilized to determine a homogenized material by calculating the material parameters $M_s, M_f, A_s, A_f, C_A, C_M, E_A, E_M, H_{sat}, k_t$. The calculation of the material parameters is repeated for three different dispersions of 15 precipitates and the calculated average material properties are considered representative of the material-at-large [52].

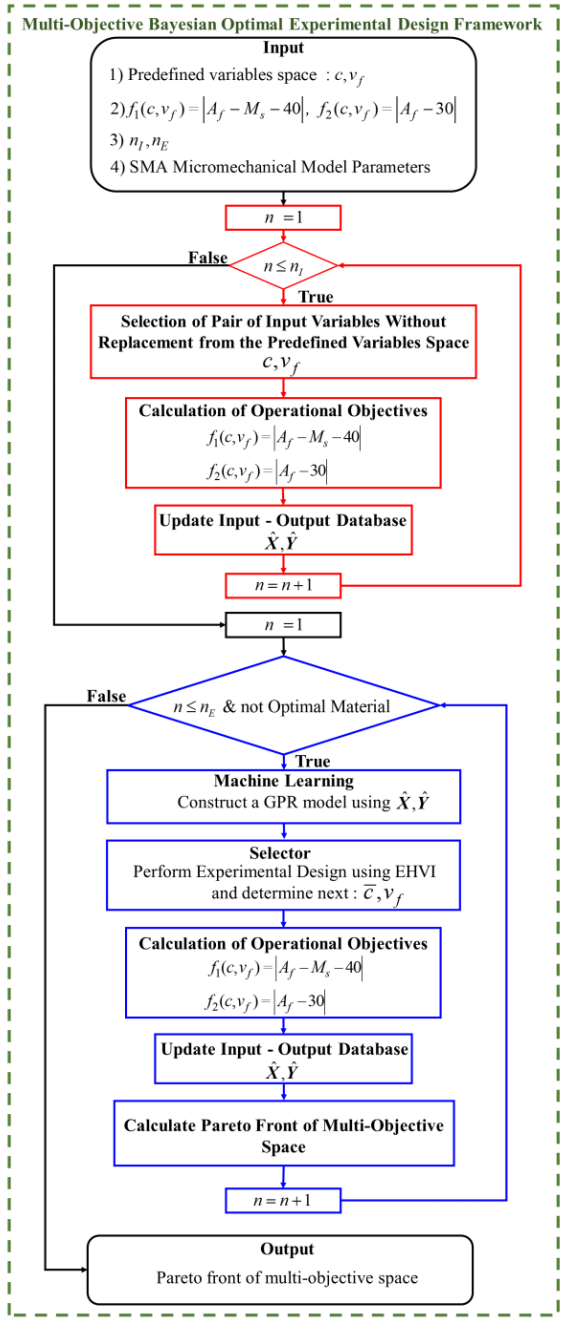

**Figure 17.** Flowchart of the multi-objective Bayesian optimal experimental design framework.

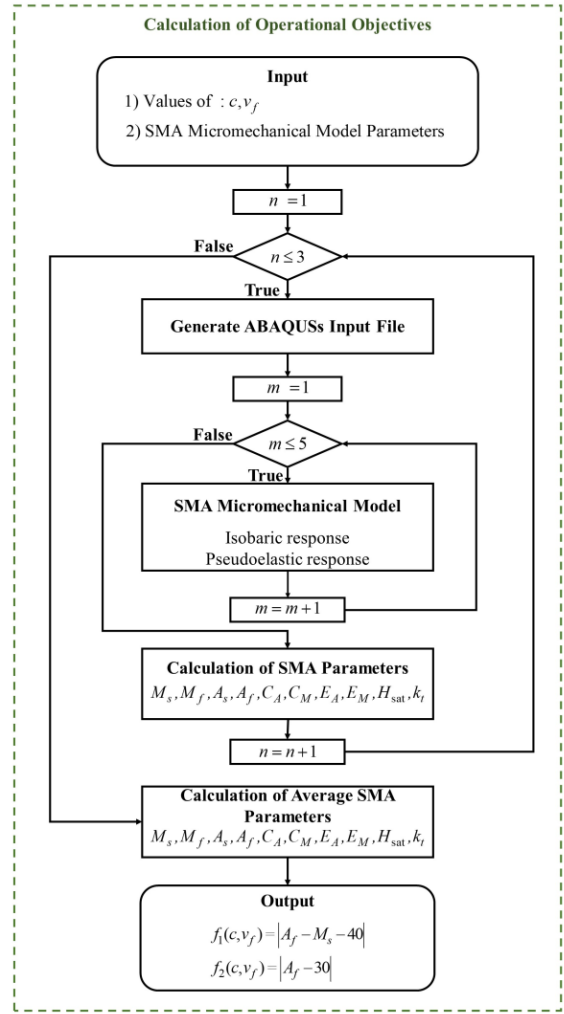

**Figure 18.** Flowchart of the calculation of the operational objectives step of the multi-objective Bayesian optimal experimental design framework.

Finally using these properties the values of the objective functions are calculated. At this point the calculation of operational objectives is completed and during the next step the initial database $\hat{X}, \hat{Y}$ of the input-output variables is updated. This loop continues until $n_I$ iterations are completed and the initial database of the input-output variables $\hat{X}, \hat{Y}$ is developed. At this point the BOED framework main iterative loop (marked with blue color in Figure 17) starts with the machine-learning step, in which the collected database is used to construct a GPR model in order to model the input-output relationships of the system under study with uncertainty. In the following selector step, the EHVI acquisition function is used to characterize the expected utility of each candidate material with respect to the GPR model and the next material to test is identified. Based on the provided inputs for the selected candidate following the process described earlier, the values of the operational objectives are computed in the updating step and the input-output database is updated. Finally the Pareto front of the measured objective space is calculated. It is worth noting that during the selector step the determined next material to test always belongs to the unexplored space due to the assumption that the results of the micromechanical model are noise free. Continuing with the description of the steps of the code, at that point if the objective functions are considered minimized then the found material satisfies all the behavior requirements and the material discovery procedure is terminated. In the case where the candidate material is not satisfactory, a new iteration of the BOED main iterative loop of the process begins and the iterations continue until the $n_E = n_B - n_I$ number of iterations is reached or a material that satisfies all the design objectives is found.

It is important to note that the developed BOED framework is general enough to be applicable beyond SMAs. To perform BOED in other material systems using the developed framework, the objective is, during the "calculation of operational objectives" step (Figure 17) to calculate the values of the defined objective functions based on the values of the design input variables. Hence the developed framework can be utilized and during the aforementioned step either physical experiments could be conducted to measure the actual response of the material and therefore to calculate the values of the defined objective functions or the equivalent computational experiments could be performed using models that simulate the behavior of the studied material as in the case presented herein. In both approaches the values of the calculated objective functions will be provided as input to the next steps of the algorithm and after the completion of each iteration of the BOED main iterative loop the next material to test will be determined.

The in-house code developed in the current work related to the BOED framework and the user's manual are shared through a GitHub repository [65].

## APPENDIX C

*Computational Efficiency of the Bayesian Optimal Experimental Design Framework*

The parallel computing capabilities of the ADA high performance research computing cluster at Texas A&M University are utilized to enhance the computational efficiency of the developed BOED framework. The SMA micromechanical model is identified to be the most computationally expensive process within the framework and therefore in order to optimally select, the number of the used physical cores to run the micromechanical simulations, the following procedure is followed.

An RVE model is generated, comprised of ~ 30,000 quadratic 10 node tetrahedral elements which represent the typical RVE model size that is used within the developed method. The RVE model is used to perform test simulations using the ABAQUS FE software [64] in order to develop the required knowledge regarding the required analysis time as a function of the used physical cores. Based on the extracted information the optimal number of physical cores is selected to perform the needed simulations. This choice leads optimal computational performance while at the same time it minimize the used number of the needed ABAQUS licensing tokens. To this end, a total number of 20 independent simulations are performed while using different number of physical cores at a time while the analysis time is measured.

Figure 19 depicts the results and it shows that the analysis time significantly reduces as the used number of physical cores increases until the initiation point of a plateau region at approximately 120 used cores.

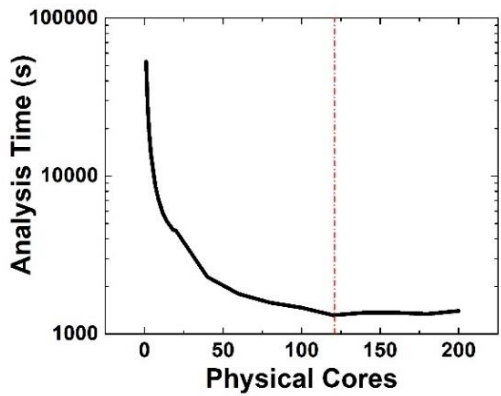

**Figure 19.** Analysis time of a typical SMA RVE model in relation to the used computer physical cores; the RVE is comprised of ~ 30,000 quadratic 10 node tetrahedral elements.

After that point, the diagram shows that no further improvement on the computational efficiency of the analysis is achieved by increasing the number of the used cores. The initiation point of the plateau region marks the number of used physical cores that maximize the computational efficiency of the analysis while the number of the used ABAQUS license tokens is minimized. In other words, the use of more than 120 cores in the conducted analysis will result to excessive use of license tokens without any reduction on the analysis time while the use of less than 120 cores will result to reduction of the computational efficiency. To this end, based on the results shown in Figure 19 in the current work, 600 physical cores are used to perform the five needed simulations per model input (precipitates volume fraction and initial Ni concenetration) in order to extract the corresponding material parameters. This choice corresponds to ~ 1300s analysis time per model input.

## DATA AVAILABILITY

The data required to reproduce these findings are available to download from the GitHub repository [65]: https://github.com/BOED-SMA.

## REFERENCES

[1] D. Jarvis, "Metallurgy Europe: A Renaissance Programme for 2012-2022.," *Adv. Phys.*, vol. 61, no. 6, pp. 665–743, 2012.

[2] G. J. Schmitz and U. Prahl, "ICMEg -- the Integrated Computational Materials Engineering expert group -- a new European coordination action," *Integr. Mater. Manuf. Innov.*, vol. 3, no. 1, p. 2, Feb. 2014.

[3] M. G. I. for G. Competitiveness, "Materials Genome Initiative for Global Competitiveness." 2011.

[4] T. Lookman, P. V. Balachandran, D. Xue, G. Pilania, T. Shearman, J. Theiler, J. E. Gubernatis, J. Hogden, K. Barros, E. BenNaim, and F. J. Alexander, "A perspective on materials informatics: State-of-the-art and challenges," in *Springer Series in Materials Science*, vol. 225, 2015, pp. 3–12.

[5] S. Curtarolo, G. L. W. Hart, M. B. Nardelli, N. Mingo, S. Sanvito, and O. Levy, "The high-throughput highway to

computational materials design," *Nat. Mater.*, vol. 12, no. 3, pp. 191–201, 2013.

[6] H. Koinuma and I. Takeuchi, "Combinatorial solid-state chemistry of inorganic materials," *Nat. Mater.*, vol. 3, no. 7, pp. 429–438, 2004.

[7] X.-D. Xiang, X. Sun, G. Briceno, Y. Lou, K.-A. Wang, H. Chang, W. G. Wallace-Freedman, S.-W. Chen, and P. G. Schultz, "A Combinatorial Approach to Materials Discovery," *Science (80-. ).*, vol. 268, no. 5218, pp. 1738–1740, 1995.

[8] S. Curtarolo, D. Morgan, K. Persson, J. Rodgers, and G. Ceder, "Predicting crystal structures with data mining of quantum calculations.," *Phys. Rev. Lett.*, vol. 91, no. 13, p. 135503, 2003.

[9] R. Y. Rubinstein and A. Shapiro, "OPTIMIZATION OF STATIC SIMULATION MODELS BY THE SCORE FUNCTION METHOD," vol. 32, pp. 373–392, 1990.

[10] P. Glasserman, *Gradient Estimation via Perturbation Analysis*. 1991.

[11] A. J. Booker, J. Dennis J.E., P. D. Frank, D. B. Serafini, V. Torczon, and M. W. Trosset, "A rigorous framework for optimization of expensive functions by surrogates," *Struct. Multidiscip. Optim.*, vol. 17, no. 1, pp. 1–13, 1999.

[12] V. Torczon, R. M. Lewis, and M. W. Trosset, "Direct search methods: then and now," *J. Comput. Appl. Math.*, vol. 124, no. 1–2, pp. 191–207, 2000.

[13] E. G. Ryan, C. C. Drovandi, J. M. Mcgree, and A. N. Pettitt, "A Review of Modern Computational Algorithms for Bayesian Optimal Design," *Int. Stat. Rev.*, vol. 84, no. 1, pp. 128–154, 2016.

[14] C. Kathryn and V. Isabella, "Bayesian Experimental Design: A Review," *Stat. Sci.*, vol. 10, no. 3, pp. 273–304, 1995.

[15] M. T. M. Emmerich, A. H. Deutz, and J. W. Klinkenberg, "Hypervolume-based expected improvement: Monotonicity properties and exact computation," *2011 IEEE Congr. Evol. Comput. CEC 2011*, pp. 2147–2154, 2011.

[16] B.-J. Yoon, X. Qian, and E. R. Dougherty, "Quantifying the objective cost of uncertainty in complex dynamical systems," *Signal Process. IEEE Trans.*, vol. 61, no. 9, pp. 2256–2266, 2013.

[17] P. Frazier, W. Powell, and S. Dayanik, "The knowledge-gradient policy for correlated normal beliefs," *INFORMS J. Comput.*, vol. 21, no. 4, pp. 599–613, 2009.

[18] W. Scott, P. Frazier, and W. Powell, "The correlated knowledge gradient for simulation optimization of continuous parameters using Gaussian process regression.," *Optim. pp*, vol. 21, no. 3 SRC-GoogleScholar FG-0, pp. 996–1026, 2011.

[19] B. Shahriari, K. Swersky, Z. Wang, R. P. Adams, and N. De Freitas, "Taking the human out of the loop: A review of Bayesian optimization," *Proc. IEEE*, vol. 104, no. 1, pp. 148–175, 2016.

[20] D. Xue, P. V. Balachandran, J. Hogden, J. Theiler, D. Xue, and T. Lookman, "Accelerated search for materials with targeted properties by adaptive design," *Nat. Commun.*, vol. 7, p. 11241, 2016.

[21] D. Xue, P. V. Balachandran, R. Yuan, T. Hu, X. Qian, E. R. Dougherty, and T. Lookman, "Accelerated search for $BaTiO_3$-based piezoelectrics with vertical morphotropic phase boundary

using Bayesian learning," *Proc. Natl. Acad. Sci.*, vol. 113, no. 47, pp. 13301–13306, 2016.

[22] R. Dehghannasiri, D. Xue, P. V. Balachandran, M. R. Yousefi, L. A. Dalton, T. Lookman, and E. R. Dougherty, "Optimal experimental design for materials discovery," *Comput. Mater. Sci.*, vol. 129, pp. 311–322, 2017.

[23] P. V. Balachandran, D. Xue, J. Theiler, J. Hogden, and T. Lookman, "Adaptive Strategies for Materials Design using Uncertainties," *Sci. Rep.*, vol. 6, no. 1, p. 19660, 2016.

[24] P. I. Frazier and J. Wang, "Bayesian optimization for materials design," *Inf. Sci. Mater. Discov. Des.*, vol. 225, pp. 45–57, 2015.

[25] A. Seko, A. Togo, H. Hayashi, K. Tsuda, L. Chaput, and I. Tanaka, "Prediction of Low-Thermal-Conductivity Compounds with First-Principles Anharmonic Lattice-Dynamics Calculations and Bayesian Optimization," *Phys. Rev. Lett.*, vol. 115, no. 20, pp. 1–5, 2015.

[26] T. Ueno, T. D. Rhone, Z. Hou, T. Mizoguchi, and K. Tsuda, "COMBO: An efficient Bayesian optimization library for materials science," *Mater. Discov.*, vol. 4, pp. 18–21, 2016.

[27] A. Seko, T. Maekawa, K. Tsuda, and I. Tanaka, "Machine learning with systematic density-functional theory calculations: Application to melting temperatures of single- and binary-component solids," *Phys. Rev. B - Condens. Matter Mater. Phys.*, vol. 89, no. 5, pp. 1–9, 2014.

[28] S. Ju, T. Shiga, L. Feng, Z. Hou, K. Tsuda, and J. Shiomi, "Designing nanostructures for phonon transport via Bayesian optimization," *Phys. Rev. X*, vol. 7, no. 2, pp. 1–10, 2017.

[29] A. Talapatra, S. Boluki, T. Duong, X. Qian, E. Dougherty, and R. Arróyave, "Efficient Experiment Design for Materials Discovery with Bayesian Model Averaging," *arXiv:1803.05460*.

[30] A. M. Gopakumar, P. V Balachandran, D. Xue, and J. E. Gubernatis, "Multi-objective Optimization for Materials Discovery via Adaptive Design," *Sci. Rep.*, no. February, pp. 1–12, 2018.

[31] D. R. Jones, M. Schonlau, and W. J. Welch, "Efficient Global Optimization of Expensive Black-Box Functions," *J. Glob. Optim.*, vol. 13, pp. 455–492, 1998.

[32] C. Edward and C. K. Williams., Rasmussen KI Williams., *Gaussian processes for machine learning*, vol. 1. MIT press Cambridge, 2006.

[33] X. Lachenal, S. Daynes, and P. M. Weaver, "Review of morphing concepts and materials for wind turbine blade applications," *Wind Energy*, vol. 16, no. 2, pp. 283–307, 2013.

[34] S. Barbarino, O. Bilgen, R. M. Ajaj, M. I. Friswell, and D. J. Inman, "A Review of Morphing Aircraft," *J. Intell. Mater. Syst. Struct.*, vol. 22, no. 9, pp. 823–877, 2011.

[35] S. Barbarino, E. I. S. Flores, R. M. Ajaj, I. Dayyani, and M. I. Friswell, "A review on shape memory alloys with applications to morphing aircraft," *Smart Mater. Struct.*, vol. 23, no. 6, 2014.

[36] D. J. Hartl and D. C. Lagoudas, "Aerospace applications of shape memory alloys," *Proc. Inst. Mech. Eng. Part G J. Aerosp. Eng.*, vol. 221, no. 4,

pp. 535–552, 2007.

[37] S. Seelecke and I. Müller, "Shape memory alloy actuators in smart structures: Modeling and simulation," *Appl. Mech. Rev.*, vol. 57, no. 1–6, pp. 23–46, 2004.

[38] L. G. Machado and M. A. Savi, "Medical applications of shape memory alloys," *Brazilian J. Med. Biol. Res.*, vol. 36, no. 6, pp. 683–691, 2003.

[39] T. Duerig, a Pelton, and D. Stöckel, "An overview of nitinol medical applications," *Mater. Sci. Eng. A*, vol. 273–275, pp. 149–160, 1999.

[40] K. Otsuka and X. Ren, "Physical metallurgy of Ti-Ni-based shape memory alloys," *Prog. Mater. Sci.*, vol. 50, no. 5, pp. 511–678, 2005.

[41] J. Michutta, C. Somsen, A. Yawny, A. Dlouhy, and G. Eggeler, "Elementary martensitic transformation processes in Ni-rich NiTi single crystals with Ni4Ti3 precipitates," *Acta Mater.*, vol. 54, no. 13, pp. 3525–3542, 2006.

[42] R. Mirzaeifar, R. DesRoches, A. Yavari, and K. Gall, "On superelastic bending of shape memory alloy beams," *Int. J. Solids Struct.*, vol. 50, no. 10, pp. 1664–1680, 2013.

[43] K. Gall, H. Sehitoglu, Y. I. Chumlyakov, I. V. Kireeva, and H. J. Maier, "The Influence of Aging on Critical Transformation Stress Levels and Martensite Start Temperatures in NiTi: Part II—Discussion of Experimental Results," *J. Eng. Mater. Technol.*, vol. 121, no. 1, p. 28, 1999.

[44] C. Calhoun, R. Wheeler, T. Baxevanis, and D. C. Lagoudas, "Actuation fatigue life prediction of shape memory alloys under the constant-stress loading condition," *Scr. Mater.*, vol. 95, no. 1, pp. 58–61, 2015.

[45] O. Karakoc, C. Hayrettin, M. Bass, S. J. Wang, D. Canadinc, J. H. Mabe, D. C. Lagoudas, and I. Karaman, "Effects of upper cycle temperature on the actuation fatigue response of NiTiHf high temperature shape memory alloys," *Acta Mater.*, vol. 138, pp. 185–197, 2017.

[46] C. Chluba, W. Ge, R. Lima de Miranda, J. Strobel, L. Kienle, E. Quandt, and M. Wuttig, "Ultralow-fatigue shape memory alloy films," *Science (80-. ).*, vol. 348, no. 6238, pp. 1004–1007, 2015.

[47] Z. Yang, W. Tirry, and D. Schryvers, "Analytical TEM investigations on concentration gradients surrounding Ni4Ti3 precipitates in Ni-Ti shape memory material," *Scr. Mater.*, vol. 52, no. 11, pp. 1129–1134, 2005.

[48] R. F. Hamilton, H. Sehitoglu, Y. Chumlyakov, and H. J. Maier, "Stress dependence of the hysteresis in single crystal NiTi alloys," *Acta Mater.*, vol. 52, no. 11, pp. 3383–3402, 2004.

[49] H. Sehitoglu, I. Karaman, R. Anderson, X. Zhang, K. Gall, H. J. Maier, and Y. Chumlyakov, "Compressive response of NiTi single crystals," *Acta Mater.*, vol. 48, no. 13, pp. 3311–3326, 2000.

[50] A. Evirgen and I. Karaman, "Microstructural characterization and shape memory response of Ni-rich NiTiHf and NiTiZr high temperature shape memory alloys," Texas A&M, 2014.

[51] A. Cox, B. Franco, S. Wang, T. Baxevanis, I. Karaman, and D. C. Lagoudas, "Predictive Modeling of the Constitutive Response of Precipitation Hardened Ni-Rich NiTi," *Shape Mem.*

*Superelasticity*, 2017.

[52] T. Baxevanis, A. Cox, and D. C. Lagoudas, "Micromechanics of precipitated near-equiatomic Ni-rich NiTi shape memory alloys," *Acta Mech.*, vol. 225, no. 4–5, pp. 1167–1185, 2014.

[53] T. Baxevanis, A. Solomou, I. Karaman, and D. C. Lagoudas, "Micromechanics and Nanomechanics of Composite Solids," in *Micromechanics and Nanomechanics of Composite Solids*, S. Meguid and G. Weng, Eds. Springer, 2018.

[54] C. Collard, T. Ben Zineb, E. Patoor, and M. O. Ben Salah, "Micromechanical analysis of precipitate effects on shape memory alloys behaviour," *Mater. Sci. Eng. A*, vol. 481–482, no. 1–2 C, pp. 366–370, 2008.

[55] C. Collard and T. Ben Zineb, "Simulation of the effect of elastic precipitates in SMA materials based on a micromechanical model," *Compos. Part B Eng.*, vol. 43, no. 6, pp. 2560–2576, 2012.

[56] J. Frenzel, E. P. George, A. Dlouhy, C. Somsen, M. F. Wagner, and G. Eggeler, "Influence of Ni on martensitic phase transformations in NiTi shape memory alloys," *Acta Mater.*, vol. 58, no. 9, pp. 3444–3458, 2010.

[57] J. Bernardini, C. Lexcellent, L. Daróczi, and D. L. Beke, "Ni diffusion in near-equiatomic Ni-Ti and Ni-Ti(-Cu) alloys," *Philos. Mag.*, vol. 83, no. 3, pp. 329–338, 2003.

[58] N. Zhou, C. Shen, M. F. X. Wagner, G. Eggeler, M. J. Mills, and Y. Wang, "Effect of $Ni_4Ti_3$ precipitation on martensitic transformation in Ti-Ni," *Acta Mater.*, vol. 58, no. 20, pp. 6685–6694, 2010.

[59] D. Lagoudas, D. Hartl, Y. Chemisky, L. MacHado, and P. Popov, "Constitutive model for the numerical analysis of phase transformation in polycrystalline shape memory alloys," *Int. J. Plast.*, vol. 32–33, pp. 155–183, 2012.

[60] M. F. X. Wagner and W. Windl, "Elastic anisotropy of $Ni_4Ti_3$ from first principles," *Scr. Mater.*, vol. 60, no. 4, pp. 207–210, 2009.

[61] D. C. Lagoudas, *Shape Memory Alloys*, vol. 1. College Station, TX, USA: Springer, 2008.

[62] E. Y. Panchenko, Y. I. Chumlyakov, I. V Kireeva, A. V Ovsyannikov, H. Sehitoglu, I. Karaman, and Y. H. J. Maier, "Effect of disperse $Ti_3N_4$ particles on the martensitic transformations in titanium nickelide single crystals," *Phys. Met. Metallogr.*, vol. 106, no. 6, pp. 577–589, 2008.

[63] A. Cox, T. Baxevanis, and D. C. Lagoudas, "Finite element analysis of precipitation effects on Ni-rich NiTi shape memory alloy response," *Mater. Sci. Forum*, vol. 792, pp. 65–71, 2014.

[64] "ABAQUS (2017) ABAQUS Documentation, Dassault Systèmes, Providence, RI, USA."

[65] G. Zhao, A. Solomou, S. Boluki, D. C. Lagoudas, R. Arróyave, and X. Qian, "Bayesian Optimal Experimental Design Framework," *GitHub Repository*, 2017. [Online]. Available: https://github.com/BOED-SMA.

[66] "MATLAB 2017, The MathWorks Inc."